\documentclass[a4paper,11pt]{article}
\pdfoutput=1 % if your are submitting a pdflatex (i.e. if you have
\usepackage{jheppub} % for details on the use of the package, please
\usepackage[T1]{fontenc} % if needed
\usepackage{comment}
\usepackage{float}
\usepackage{physics}
\usepackage{tensor}
\usepackage{ulem}
\usepackage{color}
\usepackage{mathtools}

\usepackage{hyperref}

\usepackage{graphicx}% Include figure files
\usepackage{dcolumn}% Align table columns on decimal point
\usepackage{bm}% bold math

\newcommand{\be}{\begin{equation}}
\newcommand{\ee}{\end{equation}}
\newcommand{\ba}{\begin{eqnarray}}
\newcommand{\ea}{\end{eqnarray}}

\newcommand{\beq}{\begin{equation}}
\newcommand{\eeq}{\end{equation}}
\newcommand{\beqa}{\begin{eqnarray}}
\newcommand{\eeqa}{\end{eqnarray}}

\title{\boldmath Quantum Black Holes in Non-Linear Electrodynamics}

\author[a,b]{Jana Men\v s\'iková,}
\emailAdd{jana.mensikova@matfyz.cuni.cz}
\author[a]{David Kubiz\v n\'ak,}
\emailAdd{David.Kubiznak@matfyz.cuni.cz}
\author[b]{and Robert B. Mann}
\emailAdd{rbmann@uwaterloo.ca}
\affiliation[a]{Institute of Theoretical Physics, Faculty of Mathematics and Physics,
Charles University, V Holešovičkách 2, 180 00 Prague 8, Czech Republic}
\affiliation[b]{Department of Physics and Astronomy, University of Waterloo, Waterloo, Ontario, N2L 3G1, Canada}

\abstract{Techniques of AdS/CFT correspondence and braneworld holography have proven to be a useful tool for solving the problem of semi-classical gravity where quantum matter provides the back-reaction to classical spacetime geometry. In this paper, we apply this approach to Einstein gravity coupled to RegMax nonlinear electrodynamics and construct novel type of charged quantum black holes. We then discuss their thermodynamics and horizon structure.
}

\date{\today}

\begin{document}
	\maketitle

\section{Introduction}

In the absence of a full quantum gravity theory, {\it semi-classical gravity}, e.g. \cite{Birrell:1982ix},   is a very useful approximation. It assumes that spacetime remains classical while the matter is treated as quantum, with the
energy-momentum tensor on the right-hand side of the Einstein equations  replaced with its expectation value in
some quantum state. This description allowed for the discovery of many novel and interesting phenomena, such as Hawking radiation \cite{Hawking:1975vcx}, providing the foundations for black hole thermodynamics \cite{Bardeen:1973gs, Bekenstein:1973ur}, and the famous Unruh effect \cite{Unruh:1976db}.

Standardly, when applying quantum field theory techniques in curved space, one proceeds perturbatively, quantizing first the test fields on a given background and then seeking their backreaction on the geometry. In addition to its limitations, e.g. \cite{Page:1981aj},   such a procedure is rather hard. {\it Braneworld holography} \cite{deHaro:2000wj, Emparan:2002px} offers a powerful alternative. 
Contrary to the standard AdS/CFT correspondence \cite{Maldacena:1997re, Witten:1998qj}, on the brane one has a UV finite cutoff CFT, and the gravity there remains {\it dynamical}.   
It is conjectured that the solutions of bulk gravity restricted to this brane 
encode quantum
corrections to the lower-dimensional brane geometry in all orders of backreaction \cite{Emparan:2002px, Emparan:2020znc}.
This then provides 
an effective way for  solving semi-classical gravity.

More concretely,
the idea is to insert a simple tensional brane into the AdS bulk (black hole) spacetime and expel the problematic asymptotic regions that cause divergences in the action
through means of
{\it holographic renormalization} \cite {deHaro:2000vlm}. This yields 
an induced lower-dimensional higher-curvature theory of gravity on the brane. Among its solutions are the quantum-corrected black holes (induced from the bulk) that incorporate backreaction to all orders in the matter coupling. 
Since the brane is displaced from the `center of AdS', such {\it quantum black holes} correspond to accelerated  black hole spacetimes in the bulk. Thus, by finding accelerated black hole solutions and localizing them on the brane, we automatically find solutions to the induced gravity on the brane and, hence, the corresponding quantum black holes.

In particular, 
building on the knowledge of 4-dimensional accelerated black hole spacetimes, known as the {\it C-metric}, e.g.  \cite{Cmetric, Plebanski:1976gy, Dias:2002mi, Griffiths:2005qp, Krtous:2005ej},
the braneworld construction of 3-dimensional quantum black holes \cite{Emparan:2020znc}
%Holographic renormalization 
has already been extended to charged \cite{Feng:2024uia}, \cite{Climent:2024nuj}, and rotating \cite{Emparan:2020znc,Bhattacharya:2025tdn, Cao:2026jls} cases. It provides a natural framework for describing a `higher-dimensional origin' \cite{Frassino:2022zaz} of black hole chemistry \cite{Kubiznak:2016qmn}, and was used to refine 
classical Penrose \cite{Penrose:1973um} and isoperimetric \cite{Cvetic:2010jb} inequalities into quantum settings \cite{Frassino:2024bjg}, see also \cite{Panella:2024sor} for a review. Most recently,   
a method using the quasinormal modes spectrum of a massless scalar probe has been applied to study the interiors of quantum black holes \cite{Coviello:2026tol, Coviello:2026hdc}.

Interestingly, the construction of charged quantum black holes thus far \cite{Feng:2024uia,  Climent:2024nuj} has been limited to linear Maxwell electrodynamics. 
In the (next to) leading order, this gives rise to a higher-curvature gravity coupled to linear electrodynamics on the brane. However, to treat the two theories on a `more equal footing', 
it would be more natural to consider a construction where both gravity and electrodynamics are modified on the brane already in this order.\footnote{When higher-order derivative corrections are included beyond the next to leading order, the Einstein-Maxwell theory in the bulk yields a rather complicated vector-tensor theory on the brane that, apart from gravitational higher-order derivative terms, also includes higher-order terms in vector field strength possibly coupled to spacetime curvature.}   This naturally leads us to go beyond linear Maxwell's theory in the bulk and replace it with a theory of    {\it non-linear electrodynamics (NLE)}, e.g. \cite{Born:1934gh, plebanski1970lectures, Sorokin:2021tge}.
In particular, 
there are two (4-dimensional) NLEs that are known to admit accelerated solutions, %for which such solutions are explicitly known, 
namely  the RegMax \cite{Hale:2023dpf} and the ModMax \cite{Barrientos:2022bzm} electrodynamics. 

Our goal in this paper is to focus on the former and obtain the {\it RegMax quantum black holes}; %derived from the RegMax theory; 
we shall not consider the ModMax theory 
%black holes 
in this work.  
To this end, 
we work out the counterterm structure for the RegMax NLE theory, generalizing the results of \cite{Taylor:2000xw}.
{We also obtain the thermodynamic variables for the RegMax quantum black hole and check that the first law holds.}
Moreover, apart from the thermodynamic quantities of quantum RegMax black holes, we also study their horizon structure and show that, 
unlike their Maxwellian counterparts, quantum RegMax black holes can exist in Schwarzschild-like or Reissner--Nordström-like regimes, a feature that is also typical for classical RegMax black holes \cite{Hale:2023dpf}. 

Our paper is organized as follows. We begin by  introducing the theoretical background for constructing quantum black holes on the brane in Sec.~\ref{Sec2}.  %sections \eqref{holreg}, \eqref{branes} and \eqref{qbh}. 
The main results for Maxwell quantum black holes are summarized in Sec.~\ref{qbh_maxwell}. The novel quantum black holes in RegMax non-linear electrodynamics are constructed and analyzed in Sec.~\ref{Sec4}, with  many  computational details deferred to Appendix \ref{counterterms}. Concluding remarks are in Sec.~\ref{Sec5}.

\section{Preliminaries}\label{Sec2}

\subsection{Holographic renormalization} \label{holreg}

The gravitational action in $(d+1)$-dimensional AdS spacetime reads
\be
\mathcal{I}_{grav}=\frac{1}{16\pi G_{d+1}} \int_{\cal{M}} (R-2\Lambda_{d+1})\sqrt{-g} \dd^{d+1} x + \frac{1}{8 \pi G_{d+1}}\int_{\cal{\partial M}} K \sqrt{-h} \dd^d x \,.\label{gravaction}
\ee
{Upon the variation with respect to the bulk metric $g_{\mu \nu}$,}
the first Einstein–Hilbert (or bulk) term gives the left-hand side of the Einstein field equations with  cosmological constant $\Lambda_{d+1}=-d(d-1)/2\ell^2_{d+1}$, where $\ell_{d+1}$ is the AdS radius,   and the second York–Gibbons–Hawking (or boundary) term does not contribute to the equations of motion but is necessary to ensure the Dirichlet boundary conditions.
As usual, $R$ denotes the Ricci scalar, $K$ is the extrinsic curvature, 
$G_{d+1}$ is the Newton gravitational constant, and $g$ and $h$ denote the determinants of the bulk and boundary metrics, respectively.

{When the boundary $\partial M$ is `sent to infinity',}
the action \eqref{gravaction} exhibits divergences that need to be removed. To achieve this, we use the technique of holographic renormalization \cite{deHaro:2000vlm}, which enables us to identify the divergent part of the action and add local counterterms to cancel it.
The starting point is to express the  bulk metric $g_{\mu\nu}$ near the boundary ($z=0$) in 
Fefferman--Graham coordinates {$X^\mu=(z, x^i)$ \cite{Fefferman:2007rka}
\be
g=g_{\mu \nu}\dd X^\mu \dd X^\nu=\frac{\ell^2_{d+1}}{z^2}\dd z^2+h_{ij}\dd x^i \dd x^j\,,\quad h=h_{ij}\dd x^i \dd x^j=\frac{1}{z^2}\gamma_{ij}(z,x)\dd x^i \dd x^j\,. \label{FGmetric}
\ee
}
For this metric, we calculate Einstein's equations and solve them perturbatively, order by order in $z$,  which is considered to be small. To this end, we expand the metric $\gamma_{ij}$ as
\be
\gamma_{ij}(z,x)=\gamma_{(0)ij}(x)+z^2\gamma_{(2)ij}(x)+z^4\gamma_{(4)ij}(x)
%+z^6\gamma_{(6)ij}(x)
+\dots +z^{d}\gamma_{(d)ij}(x)\,, \label{metric_expand}
\ee
where the subscripts indicate the number of derivatives of $\gamma_{(0)}$ with respect to coordinates $x^i$ 
contained in that term.
Inserting this expansion  into  Einstein's equations 
and solving them order by order in  $z$, we can   obtain all $\gamma_{(k)}$ as expressions of $\gamma_{(0)}$.

Since the bulk part of the action is divergent at the boundary $z=0$,  we   introduce a cutoff surface $z=\epsilon$ where $\epsilon$ is small. Consequently, we have the following {regularized action:}
%integration regions:
\be
\mathcal{I}_{grav}=\frac{1}{16\pi G_{d+1}} \int_{z \ge \epsilon} (R-2\Lambda_{d+1})\sqrt{-g} \dd x^{d+1} + \frac{1}{8 \pi G_{d+1}}\int_{z=\epsilon} K \sqrt{-h} \dd x^d\,, \label{cutoffaction}
\ee
where the Ricci scalar reads $R=-{d(d+1)}/{\ell^2_{d+1}}$, and the normal vector to the cutoff surface has the only nontrivial component 
$n^\rho=({z}/{\ell_{d+1}},\vec{0})$. 
  Computing the extrinsic curvature, we find that its trace is  $K=-\frac{1}{\ell_{d+1}}\left(d-\frac{z}{2} \gamma^{ij}\partial_z \gamma_{ij}\right)$. 
 Finally, we can express the determinants as follows: $\sqrt{g}=\frac{\ell^{d+1}_{d+1}}{z^{d+1}}\sqrt{\gamma}$ and $\sqrt{h}=\frac{\ell^d_{d+1}}{z^d}\sqrt{\gamma}$. The only thing we need is then to express $\sqrt{\gamma}$ as a function of $\gamma_{(0)}$ using \eqref{metric_expand}.
Computational details are rather messy and lengthy; therefore, we will not reproduce them here; they can, however, be found in \cite{Panella:2024sor} or \cite{deHaro:2000vlm}. The final result is that the gravitational action has the following divergent part: 
%can be divided into finite ($\mathcal{I}_{fin}$) and divergent ($\mathcal{I}_{div}$) parts, where
\be
\mathcal{I}_{div}=\frac{\ell_{d+1}}{16\pi G_{d+1}}\int \dd^dx\sqrt{-h}\Bigl(\frac{\textsf{R}}{d-2}+\frac{2(d-1)}{\ell_{d+1}^2}-\frac{\ell^2_{d+1}}{(d-2)^2(d-4)}\bigl(\textsf{R}^2_{ij}-\frac{d}{4(d-1)}\textsf{R}^2\bigr)+\dots\Bigr)\,, \label{divI}
\ee
where the Ricci tensor $\textsf{R}_{ij}$ and the Ricci scalar $\textsf{R}$ are computed with respect to the metric $h$, and the dots indicate higher-derivative contributions. 
Having identified the divergent part of the action, we could introduce a counterterm $\mathcal{I}_{ct}=-\mathcal{I}_{div}$ to remove the divergence (sending the boundary to infinity). 

Alternatively, instead of a cut-off surface, we can insert a brane at a fixed position near the boundary. The problematic region is then expelled from the spacetime, and {\it no divergences} are left to be removed. Because we excised part of the spacetime, we will replace it with another copy of the same AdS bulk spacetime, with the brane serving as a junction surface between the two spacetimes. 
%As we will see, 
This procedure effectively induces a higher-curvature gravity theory on the brane, where the calculated counterterms are an essential contribution.

\subsection{Braneworld holography} \label{branes}

Let us perform the task described at the end of the previous section and insert a purely tensional brane with an action
\be
\mathcal{I}_{\cal{B}}=-\tau \int_{\mathcal{B}}\sqrt{-h}\dd^dx\,, \label{Ibrane}
\ee
where $\tau$ denotes the brane tension. In doing so, we excise part of the spacetime, which we replace with another copy of the same spacetime.   Those two copies meet at the brane, which is the junction surface where the extrinsic curvature tensor is discontinuous 
\be
\Delta K_{ij}-\Delta K h_{ij}=-8\pi\tau G_{d+1} h_{ij}\,, \label{Israeli}
\ee
and is set equal to the stress energy of the brane
according to the Israel junction conditions \cite{Israel:1966rt}.
The precise location of the brane in terms of $z$ 
is governed by this condition, and in the case of a purely tensional brane, this sets the tension $\tau$.

Adding the brane action \eqref{Ibrane} to the divergent part of the bulk action \eqref{divI}, we find that a higher-curvature theory of gravity is induced on the brane with the   action 
\be
\mathcal{I}_{\mathcal{B}grav}=2\mathcal{I}_{div}+\mathcal{I}_{\mathcal{B}}=\frac{1}{16\pi G_d}\int_{\mathcal{B}} \dd^dx\sqrt{-h}\Bigl(\textsf{R}-2\Lambda_d+\frac{\ell^2_{d+1}}{(d-4)(d-2)}\Bigl(\textsf{R}^2_{ij}-\frac{d\textsf{R}^2}{4(d-1)}\Bigr){+\dots}\Bigr)\,, \label{Ibgrav}
\ee 
where we need to take twice the divergent action since we have two copies of the spacetime. As we can see, $d$-dimensional quantities have appeared in the formula satisfying \cite{Cao:2026jls}:
\be
G_d = \frac{(d-2)G_{d+1}}{2\ell_{d+1}}\,, \label{G3}
\ee
\be
\Lambda_d=-\frac{(d-1)(d-2)}{2\ell_d^2}\,, \label{Lambda3}
\ee
\be
\frac{1}{\ell_d^2}=\frac{2}{\ell_{d+1}^2}\left(1-\frac{4\pi G_{d+1}\ell_{d+1}\tau}{d-1}\right)\,.
\ee
The induced theory of gravity can, in principle, be massive, with the graviton mass depending on the position of the brane -- the closer it is to the asymptotic boundary, the smaller the mass \cite{Panella:2024sor}.

Apart from the gravitational action $\mathcal{I}_{\mathcal{B}grav}$, there is a conformal field theory (with a finite UV cutoff) living on the brane, induced holographically by the four-dimensional bulk. As per usual, we shall calculate the expectation value of its energy momentum tensor from the corresponding semi-classical Einstein equations.

\subsection{Black holes on the brane} \label{qbh}

The action \eqref{Ibgrav} derived in the previous section can be understood in two different yet equivalent ways. The first is to look at the system from the bulk perspective: \eqref{Ibgrav} describes a four-dimensional spacetime governed by Einstein gravity coupled to the brane satisfying proper boundary conditions. The second way takes the brane viewpoint: \eqref{Ibgrav} gives an action for a higher-curvature gravity theory induced on the three-dimensional brane coupled to the conformal field theory, which provides quantum backreaction on the brane metric $h_{ij}$. Since these two approaches are consistent, the solutions of bulk gravity with appropriate brane boundary conditions are exactly the same as the solutions of the semi-classical theory on the brane; in other words, the classical 4D solutions encode quantum corrections to the 3D brane geometry in all orders of backreaction.

Since black holes are amongst the solutions to gravity theories, we can naturally ask what  braneworld holography can tell us about their quantum-corrected versions. A classical black hole localized on the brane in $(d+1)$-dimensional AdS spacetime has been conjectured \cite{Emparan:2002px, Emparan:2020znc} to exactly map to a $d$-dimensional quantum-corrected black hole to  all orders of backreaction. This means that to obtain a quantum black hole, we do not have to solve (rather cumbersome) equations of motion following from the action \eqref{Ibgrav}. Instead, we can just localize the bulk black hole to the brane and the statement is that such a localized metric is a solution to induced brane gravity. Such black holes localized on the brane are accelerated with respect to the bulk \cite{Panella:2024sor}. 

Summarizing, if we want to study quantum black holes in this scenario, we have to make sure that our bulk theory of gravity (possibly coupled to some matter) admits {\it C-metric solutions}, i.e., solutions describing accelerated black holes \cite{Cmetric, Plebanski:1976gy, Dias:2002mi, Griffiths:2005qp, Krtous:2005ej}. 
While such solutions are well known in the Einstein--Maxwell theory, they may not exist (or may be currently unknown)
once we couple AdS gravity to a theory of non-linear electrodynamics.

\section{Charged quantum black holes in Maxwell electrodynamics} \label{qbh_maxwell}

\subsection{Charged matter on the brane}

The action \eqref{Ibgrav} provides an induced gravity theory on the brane in the case where there is no electric charge. If we want to study charged quantum black holes, we also have to take into consideration  the electromagnetic part of the action,  which in the case of the Maxwell theory reads
\be
\mathcal{I}_{elmag}=-\frac{1}{16\pi G_{d+1}}\int_M \dd^{d+1}x \sqrt{-g} \frac{\ell_*^2}{4}F^2+\int_{\partial M} \dd^d x \sqrt{-h}A_k j^k\,,
\ee
with {$F^2=F_{\mu\nu}F^{\mu\nu}$,} coupling length scale $\ell_*^2=\frac{16\pi G_{d+1}}{g_*^2}$, and a boundary current $j^ k$. 

Does this electromagnetic action exhibit divergences? Similar to gravity, the answer depends on the number of dimensions 
-- if we consider the standard $(3+1)$-dimensional scenario, no divergences appear. 
However, since we insert a brane at some large finite distance from the origin, we still have to calculate the counterterms (for general $d$, setting $d=3$ in the end) since those terms will contribute to the induced brane theory.

Holographic renormalization for Maxwell electrodynamics was first performed in \cite{Taylor:2000xw}, where the metric and field expansions were written
in the Fefferman--Graham form 
\ba
 &&\dd s^2=\frac{\ell_{d+1}^2}{z^2}\dd z^2+\frac{1}{z^2}\gamma_{ij}\dd x^i\dd x^j\,, \label{Marikametric}\\
 \quad  &&\gamma_{ij}=\gamma^{(0)}_{ij}+z^2\gamma^{(2)}_{ij}+z^4\gamma^{(4)}_{ij}+\dots\,,\nonumber\\
 \quad &&F=F^{(0)}+zdz\wedge A^{(1)}+z^2F^{(2)}+z^2dz\wedge A^{(2)}+z^3F^{(3)}+\dots\,,
 \label{FGexp}
 \ea
 where the forms $A^{(k)}$ and $F^{(k)}$ only depend on
 the transverse coordinates $x^i$. 
 
The divergent part of the action in the leading order is  
\ba
\mathcal{I}_{div}^{Maxwell}&=&-\frac{\ell_*^2\ell_4 (d-8) \epsilon ^{4-d}}{256 \pi  (d-4) G_{d+1}}\int \dd^d x \sqrt{-\gamma^{(0)}} F_{(0)}^2+\dots\nonumber\\
&=&-\frac{ (d-8)\ell_4 \ell_*^2}{256 \pi G_{d+1} (d-4)}\int \dd^d x \sqrt{-h} F^2+\dots\,,
\ea
where now $F^2=F_{ij}F^{ij}$\,.
The resulting divergent part in $d=3$ to the leading order is thus:
\be
\mathcal{I}_{div}^{Maxwell}=-\frac{ \ell_4 \ell_*^2}{16 \pi G_4}\int \dd^3 x \sqrt{-h} \frac{5}{16} F^2 +\mathcal{O}(\ell_4^2)\,.
\ee
Summing the divergent parts of the gravitational and electromagnetic actions with brane action gives the induced theory on the brane:
\be \label{inducedMaxwell}
\mathcal{I}_{\mathcal{B}grav}^{Maxwell}=\frac{1}{16\pi G_3}\int \dd^3 x \sqrt{-h} \Big(\textsf{R} + \frac{2}{\ell_3^2}-\frac{\Tilde{\ell}_*^2}{4}F^2+16\pi G_3 A_k j^k-\ell^2_{4}\bigl(\textsf{R}^2_{ij}-\frac{3}{8}\textsf{R}^2\bigr)+\dots\Big)\,,
\ee
where $\ell_*^2=\frac{4}{5}\Tilde{\ell}^2_*$ has been employed.
The energy-momentum tensor is then determined by a variation of this action w.r.t. the boundary metric $h_{ij}$, and reads:  
\be
8\pi G_3 \left<T_{ij}\right>=\textsf{R}_{ij}-\frac{1}{2}\textsf{R} h_{ij}-\frac{1}{\ell_3^2} h_{ij}-\frac{\Tilde{\ell}^2_*}{2}F_{ki}\tensor{F}{^k}{_j}+\frac{\Tilde{\ell}^2_*}{8}F^2 h_{ij}+\mathcal{O}(\ell_4^2) \,,\label{Tmn_maxwell}
\ee
while the electromagnetic equations of motion are
\be
\big\langle J^k\big\rangle=j^k-\frac{\Tilde{\ell^2_*}\nabla_jF^{jk}}{16\pi G_3}+\mathcal{O}(\ell_4^2)\,. \label{J_maxwell}
\ee

\subsection{Charged C-metric and its quantum black hole}
The  (electrically) charged C-metric in $(3+1)$ dimensions has the form \cite{Climent:2024nuj} 
\be\label{Cmetric_Maxwell}
\dd s^2_4=\frac{1}{\left(1+\mathcal{A}rx\right)^2}\Big(-H(r)\dd t^2+\frac{\dd r^2}{H(r)}+r^2\frac{\dd x^2}{G(x)}+r^2G(x)\dd \varphi^2\Big)\,,
\ee
where the metric functions are given by 
\ba 
H(r)&=&\frac{r^2}{\ell_3^2}+\kappa -\frac{\mu}{r\mathcal{A}}+\frac{e^2}{\mathcal{A}^2r^2}\,, \label{MaxwellH}\\
G(x)&=&1-\kappa x^2-\mu x^3-e^2x^4\,, \label{MaxwellG}
\ea
and the electrostatic potential $A$ reads 
\be\label{AMaxwell}
A=\frac{2}{\ell_*\mathcal{A}}\left(\frac{e}{r_+}-\frac{e}{r}\right)\dd t\,.
\ee
Here, $e$ denotes electric charge,  $\mu$ encodes the mass of a black hole, $\mathcal{A}$ stands for the black hole acceleration, $r_+$ is the horizon radius identified as the largest positive root of $H(r)$, $x_1$ is the smallest positive root of $G(x)$, indicating the axis of symmetry of the hole, and $\kappa$ is a discrete parameter that can take on values $1, -1$ or $0$, labeling distinct families of C-metrics; i.e., there is no invertible transformation that can bring the solution with negative $\kappa$ to the solution with positive $\kappa$.

To construct the corresponding 3-dimensional quantum black hole, we proceed as follows. First, since for the above spacetime the  surface $x=0$ is umbilic and satisfies the junction conditions, we can insert the 
tensional brane there, while `trading' the acceleration ${\cal A}$ of the bulk black hole for the brane tension $\tau$, according to  
\be \label{tauA}
{\cal A}={2\pi G_4}\tau\,.
\ee 
Second, we want to eliminate the conical singularity at $x_1$. This is achieved by 
rescaling $\varphi=\Delta \bar \varphi$, where 
\be
\Delta=\frac{2}{\abs{G'(x_1)}}=\frac{2x_1}{\abs{-3+\kappa x^2_1-e^2x^4_1}}\,,
\ee
and $\bar \varphi$ is now $(2\pi)$-periodic.
Finally, since we want to preserve the areal gauge and maintain unit lapse, we further rescale the $t$ and $r$ coordinates according to  
\be
t=\Delta\Bar{t}\,, \quad r=\frac{\Bar{r}}{\Delta}\,, %\varphi=\Delta\Bar{\varphi}\,,
\ee
and introduce a new metric function $f(\bar r)=\Delta^2H(\bar r)$.  
The metric of the three-dimensional Maxwell charged quantum black hole located on the brane is then \cite{Feng:2024uia,Climent:2024nuj}: 
\ba \label{maxwellchargedqBTZ}
\dd s^2_3&&= -f(\Bar{r})\dd \Bar{t}^2+\frac{\dd \Bar{r}^2}{f(\Bar{r})}+\Bar{r}^2\dd \Bar{\varphi}^2\,,\nonumber\\
\quad f(\Bar{r})&&=\frac{\Bar{r}^2}{\ell^2_3}-8\mathcal{G}_3 M -\frac{\mu \Delta^3}{\mathcal{A}\Bar{r}}+\frac{e^2\Delta^4}{\Bar{r}^2\mathcal{A}^2}\,,\\
\quad A&&=-\frac{2e\Delta^2}{\mathcal{A}\ell_*}\Bigl(\frac{1}{\bar r}-\frac{1}{\bar r_+}\Bigr)\dd \Bar{t}\,,\nonumber
\ea
where the 3-dimensional mass is  
\be 
M=-\frac{\kappa \Delta^2}{8\mathcal{G}_3}\,,\quad 
\mathcal{G}_3=\frac{\mathcal{A}}{2}G_4=\mathcal{A}\ell_4G_3\,,
\ee 
with ${\cal G}_3$  
being the 
renormalized Newton's constant.
Regarding the parameters of the solution, $e$ is the electric charge, $\mathcal{A}$ is now related to brane tension $\tau$ according to \eqref{tauA}, and  $\ell_3$ is the AdS\textsubscript{3} radius.  Depending on the value of $\kappa$, there exist different branches of charged quantum black holes. They are thoroughly discussed in \cite{Feng:2024uia} and \cite{Climent:2024nuj} and we will recapitulate only the main points here:
\begin{itemize}
\item In the zero backreaction limit, solutions with $\kappa=-1$ have the geometry of a black hole, whereas the  geometry with $\kappa=1,0$ corresponds to a charged defect in conical AdS\textsubscript{3} or Mink\textsubscript{3}, respectively. That means that for $\kappa=-1$ case, the backreaction only alters the already existent horizon structure. If $\kappa=1,0$, the existence of the horizon is only due to the quantum backreaction -- it dresses the charged defects.
\item Charged quantum black holes can have an extremal limit, which is present if there is a double root in the metric function. For solutions with $\kappa=-1$, the extremal limit can exist even for vanishing electric charge and the corresponding mass is then negative. On the other hand, solutions with $\kappa=1,0$ can have an extremal horizon only when charge is non-zero.
\end{itemize}

Let us now turn to thermodynamics.
Thermodynamic quantities in terms of 
\be 
z=\frac{\ell_3}{\bar r_+x_1}\,,\quad  \nu=\frac{1}{\mathcal{A}\ell_3}\,,\quad \gamma=ex_1^2\,, 
\ee 
are \cite{Feng:2024uia,Climent:2024nuj,Panella:2024sor}:
\ba
M&=&-\frac{\kappa \Delta^2}{8\mathcal{G}_3}=\frac{z^2\sqrt{1+\nu ^2} (1+\nu z) \left(1+\gamma ^2 \nu ^2 z^4+\left(\gamma ^2-1\right) \nu  z^3\right)}{2 G_3 \left(\gamma ^2 \nu ^2 z^4+2 \left(\gamma ^2+1\right) \nu  z^3+\left(\gamma ^2+3\right) z^2+1\right)^2}\,,\nonumber\\
T&=&\frac{\Delta H'(r_+)}{4\pi}=\frac{f'(\bar r_+)}{4\pi}=\frac{z \left(-\gamma ^2 \nu ^3 z^5-2 \gamma ^2 \nu ^2 z^4-\gamma ^2 \nu  z^3+\nu  z^3+3 \nu  z+2\right)}{2 \pi  \ell_3 \left(\gamma ^2 \nu ^2 z^4+2 \left(\gamma ^2+1\right) \nu  z^3+\left(\gamma ^2+3\right) z^2+1\right)}\,,\nonumber\\
S&=&\frac{2A_4}{4G_4}=\frac{\pi  r_+^2 x_1 \Delta}{G_4(1+\mathcal{A} r_+ x_1)}=\frac{\pi  \ell_3 z \sqrt{1+\nu ^2}}{G_3 \left(\gamma ^2 \nu ^2 z^4+2 \left(\gamma ^2+1\right) \nu  z^3+\left(\gamma ^2+3\right) z^2+1\right)}\,,\nonumber\\
Q&=&\frac{2}{g_*^2}\int *F=\frac{4\pi e x_1\Delta}{{\cal A} g_*^2\ell_*}=\frac{4 \sqrt{\pi} z^2  \gamma  \sqrt{1+\nu ^2} (1+\nu  z)}{\sqrt{5 G_3} g_3\left(\gamma ^2 \nu ^2 z^4+2 \left(\gamma ^2+1\right) \nu  z^3+\left(\gamma ^2+3\right) z^2+1\right)}\,,\\
\phi&=&\frac{2e\Delta^2}{{\cal A}\ell_*\bar r_+}=\frac{\sqrt{5} g_3 \gamma  \nu  z^3  (\nu  z+1)}{2\sqrt{\pi G_3} \left(2 \nu  z^3+\gamma ^2 z^2 (\nu  z+1)^2+3 z^2+1\right)}\,.\nonumber
\ea
Here, the temperature $T$ is calculated from the metric function $H(r)$ relative to the timelike Killing vector $\partial/\partial \bar t$ (it is the standard Hawking temperature for the metric \eqref{maxwellchargedqBTZ}), while the entropy $S$ is taken to be the standard entropy of the 4D black hole counted twice due to the double copy. Such an entropy 
%Note that the entropy $S$  
includes both the Wald entropy (including
contributions from higher-derivative corrections), along with the entanglement entropy of the
quantum fields.
The electric charge $Q$ is identified with the 4D electric charge, again counted twice; note 
that it vanishes in the zero backreaction limit. It means that the charge of the brane black hole is purely the effect of quantum backreaction. 
Finally, the electrostatic potential $\phi$ is found as the boundary value of the gauge field $A$, namely $\hat r\to \infty$, in a gauge where $A$ vanishes on the black hole horizon. It features 
a three-dimensional effective coupling constant  $g_3=\frac{2}{5}\frac{g_*^2}{\ell_4}$ and $\ell_*^2=\frac{4}{5}\frac{16\pi G_3}{g_3^2}$.
{Treating $z$ and $\gamma$ as independent parameters (and fixing the brane `tension' $\nu$, together with $G_3$ and $\ell_3$), one can easily verify that the above thermodynamic quantities obey the following:
\be 
\delta M=T\delta S+\phi \delta Q\,, 
\ee 
which is the standard first law.

This procedure has been routinely utilized to treat thermodynamics of quantum black holes e.g. in \cite{Emparan:2020znc, Feng:2024uia,Climent:2024nuj}. However a recent study \cite{Hennigar:2026mxv} shows that there is a subtlety that went  unrecognized before:  by inserting the brane at $x=0$, we get rid of part of the boundary, but the degree of freedom of the  remaining part  could contribute to the thermodynamics. 
This problem is remedied by the insertion of another brane at a fixed value of $r$,  which cuts off the rest of the boundary. The thermodynamics worked out above then corresponds to such a spacetime with two branes, with the black hole located on the one that fixes $x$. This situation  is illustrated in Fig.~\ref{twobranes}.

\begin{figure}[H]
\begin{center}
    \includegraphics[scale=0.40]{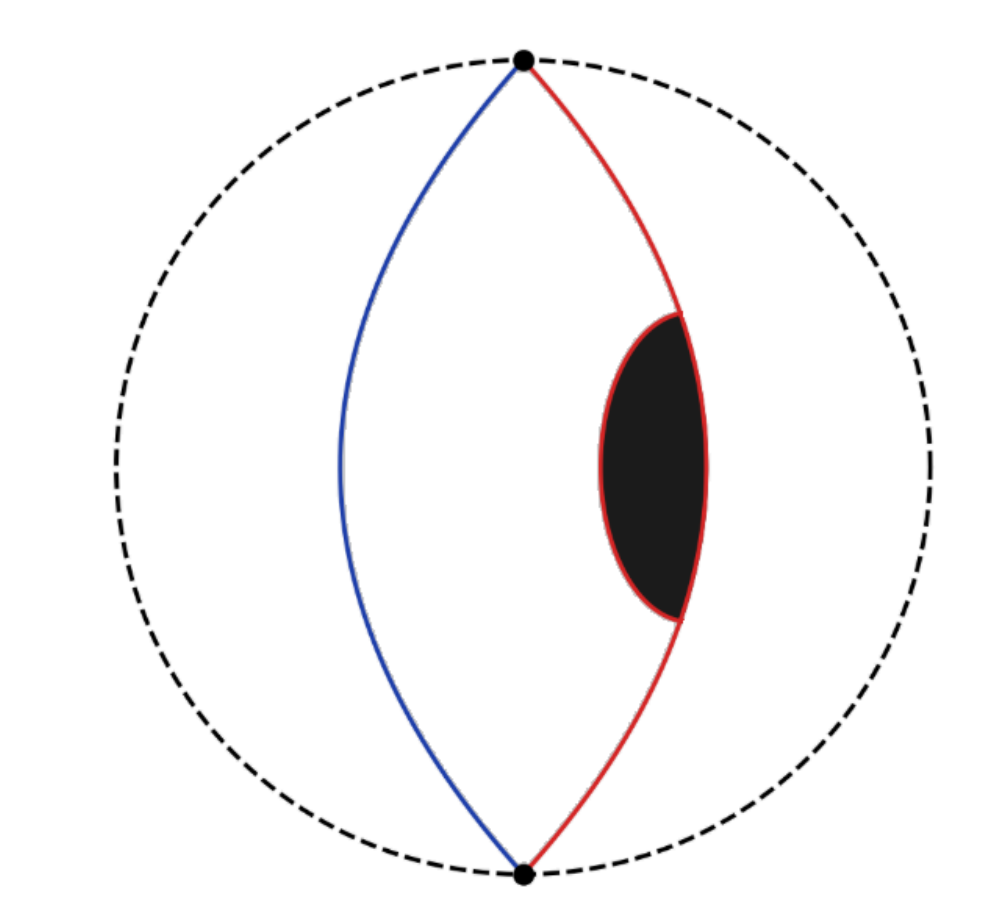}
	\caption{{\textbf{AdS spacetime with two branes inserted.} 
    In this picture, AdS spacetime is represented by a black circle with dashed boundary. At $x=0$, a brane (red curve) is inserted, which cuts off part of the boundary (to the right of the brane) but the rest of the boundary (to the left) still remains and could in principle affect the thermodynamics of the black hole localized on this brane. In order to justify the thermodynamic calculations that were done for brane black holes in this and previous papers, we have to insert another brane at $r=0$ (blue curve) which gets rid of the rest of the boundary. Subsequently, the resulting spacetime is represented by the region between the two branes; a copy of the same region can be then glued along the red brane.
        }  } 
\label{twobranes}
\end{center}
\end{figure}

We shall now consider charged black holes in which   Maxwell's theory is replaced by non-linear electrodynamics. 
 
\section{Quantum black holes in nonlinear electrodynamics}\label{Sec4}

\subsection{Nonlinear electrodynamics} \label{nle}

Theories of non-linear electrodynamics  
generalize Maxwell's theory 
by taking their Lagrangians to be `arbitrary' functions of the electromagnetic invariants to achieve the particular desired features \cite{Born:1934gh, plebanski1970lectures, Sorokin:2021tge}. In four dimensions, the two available such invariants are 
\be 
\mathcal{S}\equiv\frac{1}{2}F^2=\frac{1}{2}F_{\mu\nu}F^{\mu\nu}\,,\quad \mathcal{P}\equiv\frac{1}{2}F_{\mu\nu}   (*F)^{\mu\nu}
= \frac{1}{2}\epsilon^{\mu\nu\alpha\beta}
F_{\mu\nu}F_{\alpha\beta}\,,
\ee 
{and we have ${\cal L}={\cal L}({\cal S}, {\cal P})\,.$ Among  
such theories only two, known as the ModMax electrodynamics \cite{Bandos:2020jsw} and the RegMax electrodynamics \cite{Tahamtan:2020lvq}, are known to admit C-metric solutions, see \cite{Barrientos:2022bzm} and \cite{Hale:2023dpf}, and can be readily employed for a construction of quantum black holes.}

If we assume the action is given as 
\be 
\mathcal{I}=\frac{1}{16\pi G_4}\int \dd^4 x \sqrt{-g} \left(R+2\Lambda_4+\ell_*^2\mathcal{L}_{\mbox{\tiny NLE}}\right)\,,
\ee 
with the coupling length scale $\ell^2_*=\frac{16\pi G_4}{g^2_*}$, the energy-momentum tensor of NLE has the following form:  
\be
8\pi G_4 T_{\text{NLE}}^{\mu\nu}=-\frac{\ell_*^2}{2}\left(2F^{\mu\sigma}\tensor{F}{^\nu_\sigma}\mathcal{L}_{\mathcal{S}}+\mathcal{P}\mathcal{L}_\mathcal{P}g^{\mu\nu}-\mathcal{L}g^{\mu\nu}\right)\,, \label{eq:TNLE}
\ee
with $\mathcal{L}_{\mathcal{S}}$ and $\mathcal{L}_\mathcal{P}$ respectively denoting derivatives of the NLE Lagrangian with respect to $\mathcal{S}$ and $\mathcal{P}$. The generalized Maxwell equations now read
%field equations  also change and read
\begin{eqnarray}\label{eq:NLE equations}
dF=0\,,\quad   d*D=0\,,
\end{eqnarray}
where $D$ is an NLE field tensor
\begin{eqnarray}\label{eq:NLE tensor}
D_{\mu\nu} \equiv -2\frac{\partial{\mathcal{L}_{\mbox{\tiny{NLE}}}}}{\partial{F^{\mu\nu}}}=-2\Bigl({\cal L}_{\cal S} F_{\mu\nu}+{\cal L}_{\cal P} *F_{\mu\nu}\Bigr)\,.
\end{eqnarray}
Finally, formulas for electric and magnetic charges take the following form:
\be\label{eq:NLEcharges} 
Q_e=\frac{1}{g_*^2} \int_{S^2}*D\,,\quad Q_m=\frac{1}{g_*^2}\int_{S^2} F\,,
\ee 
where $g_*^2=\frac{5g_3^2\ell_4}{2} \approx \frac{5g_3^2}{2\mathcal{A}}$.

\subsection{RegMax theory}

For the quantum black hole construction, we consider the RegMax theory \cite{Tahamtan:2020lvq, Hale:2023dpf}, an NLE that  is unique in several aspects: {among all NLEs constructed from the invariant ${\cal S}$ in four dimensions, it is the only one that admits radiative solutions in the standard Robinson--Trautman class \cite{Tahamtan:2020lvq}, the only one that admits `Maxwell-like' slowly-rotating spacetimes \cite{Kubiznak:2022vft}, and the only one known to admit the C-metric solution \cite{Hale:2023dpf}.} Its Lagrangian reads
\be
\mathcal{L}_{\mbox{\tiny RegMax}}=-2\alpha^4\left(1-3\ln(1-s)+\frac{s^3+3s^2-4s-2}{2(1-s)}\right)\,,\quad s={\sqrt[4]{-\frac{\mathcal{S}}{\alpha^4}}} \in (0,1) \,, \label{LRM}
\ee
where the NLE parameter  $\alpha$ has dimension $[\alpha]=L^{-\frac{1}{2}}$ (the theory is no longer conformal), and the Maxwell theory is recovered in the limit $\alpha \to \infty$. Expanding in large $\alpha$, we get
\be \label{largealphaLag}
\mathcal{L}_{\mbox{\tiny RegMax}}=-\frac{1}{4}F^2-\frac{2}{5\sqrt[4]{2}\alpha}\sqrt[4]{-F^2}F^2-\frac{1}{2\sqrt{2}\alpha^2}\sqrt{-F^2} F^2+\mathcal{O}\left(\frac{1}{\alpha^3}\right)\,.
\ee
where the first term is, clearly, the Maxwell Lagrangian.
RegMax quantities like potential or field strength are convergent at the origin thanks to the following 'minimal' regularization of a field of a point charge: instead of $A=-\frac{e}{r}$ as in Maxwell's theory, we have $A=-\frac{e}{r+r_0}$ in the RegMax theory, where  $r_0=\sqrt{\abs{e}/\alpha}$, see \cite{Hale:2023dpf}. 

Gravity coupled to RegMax NLE is described by the following action:
\be
\mathcal{I}=\frac{1}{16\pi G_{4}}\int \dd^{4} x\sqrt{-g}\Bigl(R-2\Lambda_{4}+\ell_*^2\mathcal{L}_{\mbox{\tiny RegMax}}\Bigr)+\frac{1}{8\pi G_4}\int\dd^3 x \sqrt{-h} K+\int \dd^3 x \sqrt{-h}A_k j^k\,. \label{eq:RMaction}
\ee
Using \eqref{eq:TNLE}, the gravitational field equations are
\be
R_{\mu\nu}=-\frac{3}{\ell_4^2}g_{\mu\nu}+\frac{\ell_*^2 F_{\mu\sigma}\tensor{F}{_\nu ^\sigma} }{2 (s-1)^2}-\frac{g_{\mu\nu}\alpha ^4 \ell_*^2 \left(s \left(2 s^2-9 s+6\right)+6 (s-1)^2 \log (1-s)\right)}{2 (s-1)^2}\,, \label{RiccitensRM}
\ee
and have the following trace:
\be
R=-\frac{12}{\ell_4^2}-\frac{\ell_*^2}{2(s-1)}\Bigl(-\frac{2\alpha^2 s^4}{s-1} +\alpha^4 \bigl(24(s-1)\log(1-s)+12s(s-2)+4s^3\bigr)\Bigr)\,, \label{RicciRM}
\ee
with the term in the parentheses vanishing in the large $\alpha$ limit. These equations are accompanied by the generalized Maxwell equations \eqref{eq:NLE equations}, which now read $\dd F=0$ and 
\be 
\nabla_\mu\Bigl( \frac{F^{\mu\nu}}{(s-1)^2}\Bigr)=0\,.
\ee

Applying the machinery of holographic renormalization to the action \eqref{eq:RMaction}, we obtain the following action for the induced gravity on the brane:  \be \label{BgravRegMax} \mathcal{I}_{\mathcal{B}grav}=\frac{1}{16\pi G_3}\int_\mathcal{B} \dd^3 x\sqrt{-h}\left(R+\frac{2}{\ell_3^2}+\ell_4^2\left(\frac{3}{8}R^2-R_{ij}R^{ij}\right)+\dots\right)+\mathcal{I}_{EM}
 \ee
 with 
\ba\label{IEMmain}
\mathcal{I}_{EM}&=&\frac{\Tilde{\ell}^2_*}{16 \pi G_3}\int \dd^3 x \sqrt{-h}\left(-\frac{1}{4} F^2-\frac{6}{25 \sqrt[4]{2}  \alpha}\sqrt[4]{-F^2} F^2 +\frac{1}{5 \sqrt{2}  \alpha ^2}\sqrt{-F^2} F^2+\dots\right)\nonumber\\
&&+2\int \dd^3 x \sqrt{-h}A_k j^k\,,
\ea
and $\Tilde{\ell}^2_*=\frac{5}{4}\ell_*^2$. Note that the dots stand for terms in higher orders of $\alpha$ , c.f. (four-dimensional)  \eqref{largealphaLag}. Notice that while the structure is the same, the numerical coefficients differ, implying that the NLE theory localized on the brane is different from the bulk NLE. 
The derivation of the electromagnetic counterterms is rather lengthy and tedious, and we provide the step-by-step calculation based on the approach in \cite{Taylor:2000xw} in  Appendix \ref{counterterms}. 

 To derive the energy-momentum tensor $\left<T_{ij}\right>$, we have to vary the action $\mathcal{I}_{\mathcal{B}grav}$ with respect to the metric $h$. Using 
\be
\delta \sqrt{-h}=-\frac{1}{2}h_{ij}\sqrt{-h}\delta h^{ij}\,,
\ee
together with 
\ba
\frac{\delta F^{kl}}{\delta h^{ij}}&=&\frac{\delta \left(h^{km}h^{ln}F_{mn}\right)}{\delta h^{ij}}=F_{mn}\left(h^{km}\delta^l_i\delta^n_j+h^{ln}\delta^k_i\delta^m_j\right)=F_{mj}h^{km}\delta^l_i+F_{jn}h^{ln}\delta^k_i\,,\nonumber\\
\frac{\delta F^2}{\delta h^{ij}}&=&\frac{F_{kl}\delta F^{kl}}{\delta h^{ij}}=F_{kl}\left(F_{mj}h^{km}\delta^l_i+F_{jn}h^{ln}\delta^k_i\right)=F_{ki}\tensor{F}{^k}{_j}+F_{il}\tensor{F}{_j}{^l}=2F_{ki}\tensor{F}{^k}{_j}\,,\nonumber\\
\frac{\delta \sqrt{-F^2}}{\delta h^{ij}}&=&-\frac{1}{2 \sqrt{-F^2}}\frac{\delta F^2}{\delta h^{ij}}=-\frac{F_{ki}\tensor{F}{^k}{_j}}{\sqrt{-F^2}}\,,\\
\frac{\delta \sqrt[4]{-F^2}}{\delta h^{ij}}&=&-\frac{1}{4 (-F^2)^{3/4}}\frac{\delta F^2}{\delta h^{ij}}=-\frac{F_{ki}\tensor{F}{^k}{_j}}{2(-F^2)^{3/4}}\,,\nonumber
\ea
 we find 
\ba
8\pi G_3 \left<T_{ij}\right>&&=\textsf{R}_{ij}-\frac{1}{2}\textsf{R} h_{ij}+\Lambda_3 h_{ij}-\frac{\Tilde{\ell}^2_*}{2}F_{ki}\tensor{F}{^k}{_j}+\frac{\Tilde{\ell}^2_*}{8}F^2 h_{ij}+\frac{3\Tilde{\ell}^2_*}{25 \sqrt[4]{2}  \alpha}\sqrt[4]{-F^2} F^2 h_{ij}\nonumber\\
&&\quad -\frac{\Tilde{\ell}^2_*}{10 \sqrt{2}  \alpha ^2}\sqrt{-F^2} F^2 h_{ij}-\frac{12\Tilde{\ell}^2_*}{25 \sqrt[4]{2}  \alpha}\sqrt[4]{-F^2}F_{ki}\tensor{F}{^k}{_j}+\frac{3\Tilde{\ell}^2_*}{25 \sqrt[4]{2}  \alpha}\frac{F_{ki}\tensor{F}{^k}{_j}}{(-F^2)^{3/4}}F^2\nonumber\\
&&\quad +\frac{2\Tilde{\ell}^2_*}{5 \sqrt{2}  \alpha ^2}\sqrt{-F^2} F_{ki}\tensor{F}{^k}{_j}-\frac{\Tilde{\ell}^2_*}{5 \sqrt{2}  \alpha ^2}F^2\frac{F_{ki}\tensor{F}{^k}{_j}}{\sqrt{-F^2}} + 16\pi G_3 A_k j^k h_{ij}+\dots\nonumber\\
\quad
\ea
for the CFT energy-momentum tensor. 
In the limit   $\alpha \to \infty$, it reduces to \eqref{Tmn_maxwell}.

We now consider variation with respect to the electromagnetic field.  
Using
\be
\frac{\delta F^2}{\delta \partial_i A_j} =4F^{ij}\,,
\quad 
\frac{\delta \sqrt{-F^2}}{\delta \partial_i A_j} =-\frac{2F^{ij}}{\sqrt{-F^2}}\,,
\quad 
\frac{\delta \sqrt[4]{-F^2}}{\delta \partial_i A_j}  =-\frac{F^{ij}}{(-F^2)^{3/4}}\,,
\ee
we then find
\ba
\big\langle J^k\big\rangle&&=j^k-\frac{\Tilde{\ell^2_*}\nabla_jF^{jk}}{16\pi G_3}+\frac{3\Tilde{\ell^2_*}}{50 \pi G_3 \sqrt[4]{2}\alpha}\nabla_j\left(F^{jk}\sqrt[4]{-F^2}\right)-\frac{3\Tilde{\ell^2_*}}{200 \pi G_3 \sqrt[4]{2}\alpha}\nabla_j\left(\frac{F^{jk}F^2}{(-F^2)^{3/4}}\right)\nonumber\\
&&\quad +\frac{\Tilde{\ell^2_*}}{20 \pi G_3 \sqrt{2}\alpha^2}\nabla_j\left(F^{jk}\sqrt{-F^2}\right)-\frac{\Tilde{\ell^2_*}}{40 \pi G_3 \sqrt{2} \alpha^2}\nabla_j \left(\frac{F^2F^{jk}}{\sqrt{-F^2}}\right)+\dots\,,
\ea
where the first term comes from  the boundary term $\int \dd^3 x \sqrt{-h} j^k A_k$, and the electromagnetic current $j^k$ can be obtained from the NLE electromagnetic junction conditions.
This again reduces to the Maxwell expression \eqref{J_maxwell} in the limit of large $\alpha$.

\subsection{RegMax quantum black hole} \label{qbh_regmax}
The equations of motion following from \eqref{eq:RMaction} admit an
AdS C-metric solution. If we perform rescaling 
$\alpha^4\ell_*^2 \to {\alpha}^4,~\ell_*^2\mathcal{S} \to {\mathcal{S}}$\,,
to absorb the coupling length-scale $\ell_*^2$,  
making effectively both $\alpha$ and ${\cal S}$ dimensionless, the C-metric is given by \cite{Hale:2023dpf} 

\ba\label{eq:RegMaxCmetric}
\dd s^2_4&=&\frac{1}{(x+y)^2}\Bigl(-\Tilde{H}(y)\dd t^2+\frac{\dd y^2}{\Tilde{H}(y)}+\frac{\dd x^2}{\Tilde{G}(x)}+\Tilde{G}(x)\dd\varphi^2\Bigr)\,, \nonumber\\
A&=&-\frac{\alpha e y}{\alpha+y\sqrt{|e|}}\dd t\,,
\ea 
where 
\ba 
\Tilde{H}&=&-4\alpha^4\log\Bigl(1+\frac{\sqrt{|e|}y}{\alpha}\Bigr)+\Bigl(\frac{c_1}{6}+\frac{4|e|^{3/2}\alpha}{3}\Bigr)y^3-\Bigl(\frac{c_2}{2}+2|e|\alpha^2\Bigr)y^2\nonumber\\
&&+\Bigl(c_3+4\sqrt{|e|}\alpha^3\Bigr)y-c_4+\frac{1}{\ell_4^2}\,,\nonumber\\
\Tilde{G}&=&4\alpha^4\log\Bigl(1-\frac{\sqrt{|e|}x}{\alpha}\Bigr)+\Bigl(\frac{c_1}{6}+\frac{4|e|^{3/2}\alpha}{3}\Bigr)x^3+
\Bigl(\frac{c_2}{2}+2|e|\alpha^2\Bigr)x^2\nonumber\\
&&+\Bigl(c_3+4\sqrt{|e|}\alpha^3\Bigr)x+c_4\,,
\ea
and $c_1,\dots,c_4$ are arbitrary integration constants to be specified below. 

Let us first bring this metric into a form similar to \eqref{Cmetric_Maxwell}--\eqref{AMaxwell}.
Performing the  following coordinate transformations %\tcr{(I will try it without $\omega$)}
\be\label{eq:trafo1}
y\to\frac{1}{r}\,,\quad x\to {\cal A} x\,,\quad {\varphi\to \frac{\varphi}{{\cal A}}}\,, 
\ee
\be\label{eq:trafo2} 
\Omega=r (x+y)\,,\quad H=r^2\Tilde{H}\,,\quad G=\frac{\Tilde{G}}{{\cal A}^2}\,,
\ee 
where $\mathcal{A}$ denotes acceleration, we obtain 
\ba\label{RegMax solution}
\dd s^2_4&=&g_{\mu\nu}\dd x^\mu \dd x^\nu=\frac{1}{\Omega^2}\Bigl(-H(r)\dd t^2+\frac{\dd r^2}{H(r)}+r^2\Bigl[\frac{\dd x^2}{G(x)}+ G(x) \dd\varphi^2\Bigr]\Bigr)\,,\nonumber\\ 
A&=&\Bigl(-\frac{\alpha e}{\alpha r+\sqrt{|e|}}+\frac{\alpha e}{\alpha r_++\sqrt{|e|}}\Bigr)\dd t\,,\quad 
\Omega=1+{\cal A}rx\,.
\ea
Moreover, we can now choose the values of the integration constants. Let us set $c_3=0$ and $c_4=\mathcal{A}^2$ to ensure that $x=0$ is an umbilic surface with  extrinsic curvature   $K_{ij}=-\mathcal{A}h_{ij}$.
%which fixes the tension of the brane inserted at $x=0$ to $\tau=\frac{\mathcal{A}}{2\pi G_4}$. 
Additionally, let us choose $c_1=-\frac{6\mu}{\mathcal{A}}$ and $c_2=-2\kappa$, while also rescaling the electric charge as $e\to \frac{e}{\mathcal{A}}$; we shall also, in what follows, assume, w.l.o.g., $e$ to be positive. This leads to the following metric functions:
\ba \label{RMmetricfunctions}
H&=&\kappa-\mathcal{A}^2 r^2-4 \alpha ^4 r^2 \log \Bigl(\frac{\sqrt{\frac{e}{\mathcal{A}}}}{\alpha  r}+1\Bigr)+4 \alpha ^3 r \sqrt{\frac{e}{\mathcal{A}}}+\frac{4 \alpha  e \sqrt{\frac{e}{\mathcal{A}}}-6 \alpha ^2 e r-3 \mu }{3 \mathcal{A} r} +\frac{r^2}{\ell_4^2}\,,\nonumber\\
G&=&1-\kappa  x^2-\mu  x^3+\frac{4 \alpha ^4}{{\cal A}^2} \log\Bigl(1-\frac{x \sqrt{{e}{\mathcal{A}}}}{\alpha }\Bigr)+\frac{4}{3} \alpha  e x^3 \sqrt{\frac{e}{\mathcal{A}}}+\frac{2 \alpha ^2 e x^2}{\mathcal{A}}+\frac{4 \alpha ^3 x \sqrt{\frac{e}{\mathcal{A}}}}{\mathcal{A}}\,.\qquad\ \ 
\ea
Provided we also use 
\be\label{l4} 
\ell_4^2=\left(\mathcal{A}^2+\frac{1}{\ell^2_3}\right)^{-1}\,,
\ee 
these metric functions reproduce exactly \eqref{MaxwellH} and \eqref{MaxwellG} in the large $\alpha$ limit.  

To construct the corresponding 3d quantum black hole, we now proceed as in the Maxwell case. First, we insert the brane at the umbilic $x=0$ surface, `binding' the black hole  acceleration ${\cal A}$ to the brane tension $\tau$, according to \eqref{tauA}.
Second, to ensure that we have smooth geometry along $x=x_1$, 
we introduce a new $(2\pi)$-periodic 
 $\bar \varphi$ according to   $\varphi=\Delta \bar \varphi$, where
\ba\label{rmdelta}
\Delta&=&\frac{2}{\abs{G'(x_1)}}\,,
\ea 
 and 
\ba 
G'(x_1)&=& -\frac{3}{x_1}-\frac{12\alpha^4}{{\cal A}^2 x_1}  \log \Bigl(1-\frac{\mathcal{A} x_1 \sqrt{\frac{e}{\mathcal{A}}}}{\alpha }\Bigr)+\frac{2 \alpha ^2 e x_1 }{\mathcal{A}}\left(\frac{2 \alpha ^2}{\mathcal{A} e x_1^2-\alpha ^2}-1\right)
\nonumber\\
&&+\frac{4 \alpha ^3 \sqrt{\frac{e}{A}} \left(2 \mathcal{A} e x_1^2-3 \alpha ^2\right)}{\mathcal{A}\left(\alpha ^2-\mathcal{A} e x_1^2\right)}\,.
\ea 
Finally, since we want to preserve the areal gauge and maintain
unit lapse, we further rescale the $t$ and $r$ coordinates according to
\be
t=\Delta\Bar{t}\,, \quad r=\frac{\Bar{r}}{\Delta}\,, %\varphi=\Delta\Bar{\varphi}\,,
\ee
and introduce a new metric function $f_{\mbox{\tiny qRM}}(\bar r)=\Delta^2H(\bar r)$. This yields the following RegMax quantum black hole:
\ba \label{qRegMax solution}
\dd {s}^2_3&=&
%h_{\mu\nu}\dd x^\mu \dd x^\nu=
-{f}_{\mbox{\tiny qRM}}\dd \Bar{t}^2+\frac{\dd \Bar{r}^2}{{f}_{\mbox{\tiny qRM}}}+\Bar{r}^2 \dd\Bar{\varphi}^2\,,\nonumber\\
A&=&
\frac{\alpha e \Delta^2}{\cal A}\Bigl(-\frac{1}{ \Delta \sqrt{\frac{e}{\mathcal{A}}}+\alpha  \Bar{r}}+\frac{1}{ \Delta \sqrt{\frac{e}{\mathcal{A}}}+\alpha  \Bar{r}_+}\Bigr)\dd \Bar{t}\,,\quad \,.
\ea
where
\be 
{f}_{\mbox{\tiny qRM}}(\bar r)=\Delta^2\kappa-\frac{\mu\Delta^3 }{\mathcal{A} \bar r}+\frac{\bar r^2}{\ell_3^2}-\frac{2 \alpha ^2 \Delta^2 e}{\mathcal{A}}-4 \alpha ^4 \bar r^2 \log \bigg(\frac{\sqrt{\frac{e}{\mathcal{A}}}\Delta}{\alpha  \bar r}+1\bigg)+4 \alpha ^3 \Delta \bar r \sqrt{\frac{e}{\mathcal{A}}}+\frac{4 \alpha  \Delta^3\left(\frac{e}{\mathcal{A}}\right)^{3/2}}{3 \bar r}\,. 
\ee

The  metric \eqref{qRegMax solution} is a solution to the induced gravity theory \eqref{BgravRegMax} on the brane. In accordance with \cite{Climent:2024nuj} or \cite{Emparan:2020znc}, we now identify the mass with the absolute term of the large $\bar r$ expansion of ${f}_{\mbox{\tiny qRM}}$. Thus we have 
\be \label{Mf}
{f}_{\mbox{\tiny qRM}} = \frac{\bar r^2}{\ell_3^2}+\Delta ^2 \kappa -\frac{\mu \Delta^3}{{\cal A}\bar r}+\mathcal{O}\left(\frac{1}{{\bar r}^2}\right)\equiv 
\frac{\bar r^2}{\ell_3^2}-8\mathcal{G}_3M -\frac{\mu \Delta^3}{{\cal A}\bar r}+\mathcal{O}\left(\frac{1}{{\bar r}^2}\right)
\,,
\ee 
%or more precisely, this term is equal to $-8\mathcal{G}_3M$. 
where 
$\mathcal{G}_3$ denotes the renormalized Newton's constant,  given by
\be
\mathcal{G}_3=\ell_4\mathcal{A} G_3=\frac{\mathcal{A} G_4}{2} \,.
\ee
It accounts for the effects of higher-curvature terms of the induced brane gravity on three-dimensional mass. This identification leads to
\be
M=-\frac{\sqrt{1+\mathcal{A}^2\ell_3^2}}{2\ell_3 G_3}\frac{\mathcal{A}^3 \kappa  x_1^2 \Bigl(\alpha ^2+\mathcal{A} x_1 \left(e x_1-2 \alpha  \sqrt{\frac{e}{\mathcal{A}}}\right)\Bigr)}{ \textrm{denom}_M}\,, \label{qRMmass}
\ee
where 
\ba
\textrm{denom}_M&&=\mathcal{A} x_1 \sqrt{\frac{e}{\mathcal{A}}} \Bigl(\mathcal{A} x_1^2 \left(\mathcal{A} \kappa -2 \alpha ^2 e\right)-3\mathcal{A}^2+12\alpha ^4\Bigr)-\alpha \mathcal{A} \Bigl(x_1^2 \left(\mathcal{A} \kappa +6 \alpha ^2 e\right)-3 \mathcal{A}\Bigr)\nonumber\\
&&\quad +12 \alpha ^4 \Bigl(\alpha -\mathcal{A} x_1 \sqrt{\frac{e}{\mathcal{A}}}\Bigr) \log \Bigl(1-\frac{\mathcal{A} x_1 \sqrt{\frac{e}{\mathcal{A}}}}{\alpha }\Bigr)\,.
\ea
%\ba
%\textrm{denom}_M&&=\mathcal{A} x_1 \sqrt{\frac{e}{\mathcal{A}}} \Bigl(\mathcal{A} x_1^2 \left(\mathcal{A} \kappa -2 \alpha ^2 e\right)-3 \left(\mathcal{A}^2-4 \alpha ^4\right)\Bigr)-\alpha \mathcal{A} \Bigl(x_1^2 \left(\mathcal{A} \kappa +6 \alpha ^2 e\right)-3 \mathcal{A}\Bigr)\nonumber\\
%&&\quad +12 \alpha ^4 \Bigl(\alpha -\mathcal{A} x_1 \sqrt{\frac{e}{\mathcal{A}}}\Bigr) \log \Bigl(1-\frac{\mathcal{A} x_1 \sqrt{\frac{e}{\mathcal{A}}}}{\alpha }\Bigr)\,.
%\ea

\subsection{Horizon structure}

%Having our 
With the 
metric  \eqref{qRegMax solution} for the quantum RegMax black hole in hand, we can now turn to studying its horizon structure. Classical 4D (non-accelerating) RegMax black holes are known to have two different possible branches depending on the value of the black hole mass $M$ (and its charge $Q$) 
%NLE parameter $\alpha$ 
\cite{Hale:2023dpf}.
Namely, for 
\be 
M>M_c\equiv\frac{2}{3}\alpha |Q|^{3/2}\,,
\ee 
one has a {\it Schwarzschild-like} branch with one non-degenerate horizon, while for $M<M_c$ one has a Reissner--Nordström-like branch with two horizons, one degenerate horizon, or no horizon; the case $M=M_c$ is marginal between the two.

Let us now study the situation for RegMax quantum black holes. We can perform a small $\bar r$ expansion of the metric function $ f_{\mbox{\tiny qRM}}$ to obtain
\be
f_{\mbox{\tiny qRM}} = \frac{\Delta^3}{\mathcal{A}\bar r}\left(\frac{4}{3}\alpha  e \sqrt{\frac{e}{\cal A}}-\mu\right)+\Delta^2\left(\kappa-\frac{2\alpha^2e}{{\cal A}}\right)+\mathcal{O}(\bar r)\,.
\ee
From this, we can infer that for 
\be 
\mu>\mu_c\equiv \frac{4}{3}\alpha  e \sqrt{\frac{e}{\cal A}}\,,
\ee 
we have the Schwarzschild-like branch, and for $\mu<\mu_c$ we have the Reissner–Nordström-like branch. Note, however, that contrary to the 4D classical RegMax case, $\mu$ is no longer straightforwardly related to the mass of the hole (which is now determined from the $\kappa \Delta^2$ term); rather, $\mu$ is a function of $x_1, e$ and $\kappa$, expressed from $G(x_1)=0$ in \eqref{RMmetricfunctions}.
The marginal case occurs for $\mu=\mu_c$; when $\kappa=1$ we can also find a value of $\alpha$ such that the absolute term is zero, which would give us a marginal case with the event horizon in $\bar r = 0$.
We display the corresponding horizon structure for $\kappa=1$ in Fig.~\ref{qRMhorizons1}; $\kappa=0$ and $\kappa=-1$ behave qualitatively similarly, and we do not display them here.

\begin{figure}[H]
\begin{center}
    \includegraphics[scale=0.82]{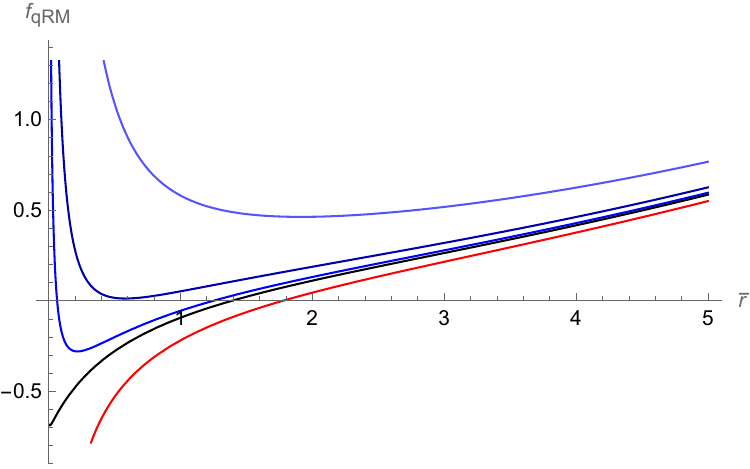}
	\caption{{\textbf{Horizon structure of quantum RegMax black hole for $\kappa=1$.} This plot shows a horizon structure for a quantum RegMax black hole with $\kappa=1, ~\mathcal{A}=0.2, ~\ell_3=8, ~e=0.5$, $\alpha=0.8$ and varying $\mu$. There are two distinct regimes: Schwarzschild-like (red) branch for $\mu>\mu_c\approx 0.84327$ with one non-degenerate horizon (displayed for $\mu=1$) and Reissner--Nordström-like (blue) branch for $\mu<\mu_c$ with two horizons ($\mu=0.8$), one degenerate horizon ($\mu \approx 0.69$), or no horizons ($\mu=0.3$). The two are divided by a black curve marginal case ($\mu=\mu_c$). The horizon structure for $\kappa=0$ and $\kappa=-1$ cases is  qualitatively similar.}}
\label{qRMhorizons1}
\end{center}
\end{figure}

\begin{figure}[H]
\begin{center}
    \includegraphics[scale=0.82]{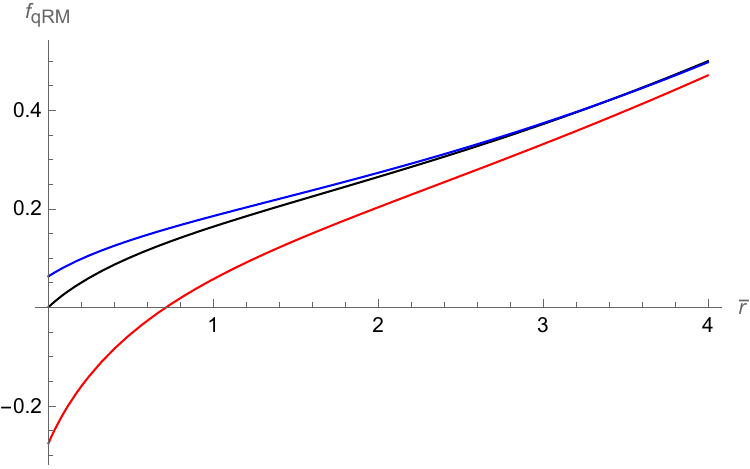}
	\caption{{\textbf{Quantum RegMax black hole: marginal case.} In this plot, we can observe marginal curves for $\kappa=1$, $\mathcal{A}=0.2$, $e=0.5$ and $\ell_3=8$. Marginal curves dividing Schwarzschild-like and Reissner--Nordström-like branches move upward with decreasing $\alpha$ and when they get above the $\bar r$-axis, they prevent the existence of extremal and two-horizon black holes. Marginal curves for $\kappa =-1$ or $\kappa=0$ cannot get so high thus for them, both RN-like and Schwarzschild-like black holes are always possible. Values of $\alpha$ used in this figure are $\alpha\in\{0.6, 0.447, 0.4\}$ for the red, black, and blue curves, respectively.}}
\label{marginal}
\end{center}
\end{figure}

In Fig.~\ref{marginal}, we take a closer look at the marginal case for $\kappa=1$. With decreasing $\alpha$, the marginal curves move up until they reach $\bar r=0$. From here on, it is not possible to have two horizons or one extremal horizon -- the Reissner–Nordström--like branch reduces to naked singularities, and the only possible black holes are of Schwarzschild-like type. When $\kappa=-1,0$, the marginal curve cannot move all the way up to $\bar r=0$, and the full Reissner--Nordström-like branch is preserved.

The above discussed horizon structure of quantum RegMax BTZ black holes is  very different from that of classical charged BTZ \cite{Banados:1992wn}
%, maxwellchargedqBTZ} 
and quantum Maxwell \cite{Feng:2024uia, Climent:2024nuj} black holes. 
The former is given by \cite{Banados:1992wn}: %maxwellchargedqBTZ}:
\ba
\dd s^2_{\mbox{\tiny BTZ}}&=&-f_{\mbox{\tiny BTZ}}\dd t^2+\frac{\dd r^2}{f_{\mbox{\tiny BTZ}}}+r^2\dd \varphi^2\,,\nonumber\\
A&=&e\log(r/r_0)\dd t\,,\quad 
f_{\mbox{\tiny BTZ}}=\frac{r^2}{\ell^2_3}-m-2e^2\log(r/r_0)\,,
\ea
where $r_0$ is an arbitrary length scale, 
and the latter was given in~\eqref{maxwellchargedqBTZ}.
In both cases, the corresponding metric functions only feature the Reissner--Nordström-type branches, as displayed in Fig.~\ref{BTZclass} for the classical BTZ case and in Fig.~\ref{BTZquantum} for the quantum Maxwell case. In the classical RegMax case, the existence of the Schwarzschild-like branch can be related to the finite field strength of point charges, e.g. \cite{Hale:2025ezt, Russo:2026vnj}. This property then seems to induce the existence of such a branch for the quantum RegMax case; though, as described above, the direct relationship to the (electromagnetic) mass of the hole is no longer straightforward.\footnote{In fact, a horizon structure similar to what we observe here for quantum RegMax black holes was also observed for classical 3D  solutions of regularized conformal electrodynamics, see \cite{Kubiznak:2024ijq}.}

\begin{figure}[H]
\begin{center}
    \includegraphics[scale=0.67]{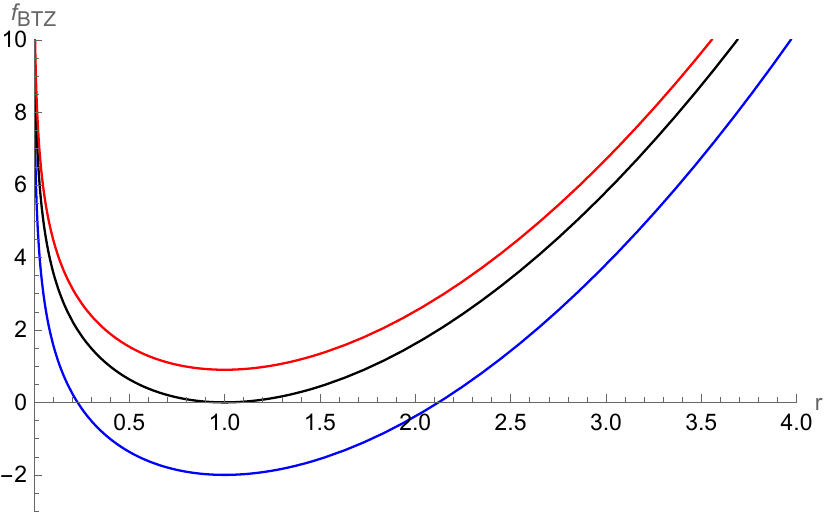}
	\caption{{\textbf{Horizon structure of a classical charged BTZ black hole.} In this plot, we can see a horizon structure of a classical charged BTZ black hole with $e=1,\ell_3=1$. Depending on the value of the mass $m$, there can be two horizons (blue, $m=3$, one extremal horizon (black, $m \approx 1$) or naked singularity (red, $m=0.1$). There will never appear a Schwarzschild-like branch. }}
\label{BTZclass}
\end{center}
\end{figure}

\begin{figure}[H]
\begin{center}
    \includegraphics[scale=0.67]{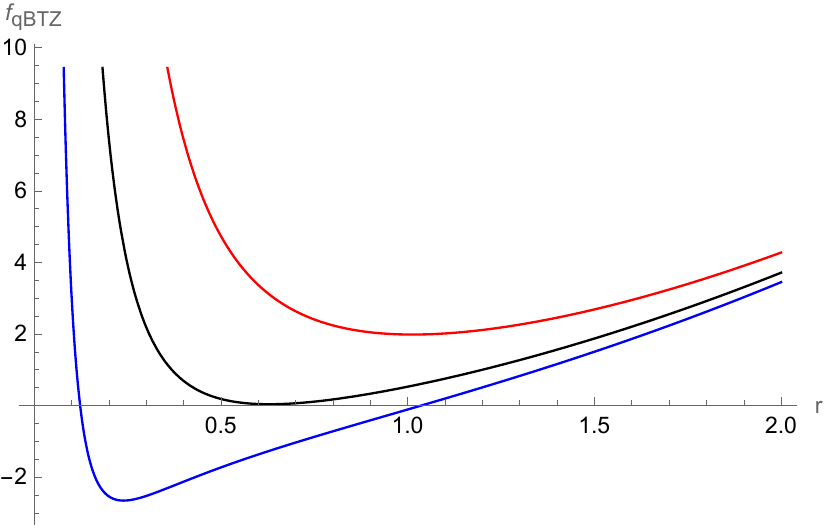}
	\caption{{\textbf{Horizon structure of quantum Maxwell BTZ black hole.} The horizon structure of a qunatum charged BTZ black hole is illustrated here with $\kappa=1,\ell_3=1,e=0.9$ and $\mathcal{A}=0.2$. Varying the mass parameter $\mu$, there can be two horizons (blue, $\mu=10$, one extremal horizon (black, $\mu \approx 3.6$) or naked singularity (red, $\mu=1$). Again, there is no Schwarzschild-like branch. }}
\label{BTZquantum}
\end{center}
\end{figure}

Let us finally turn towards the {\it extremal} quantum RegMax black holes. The corresponding extremal 
parameters $\mu_{ext}$ and $e_{ext}$ can be found from the condition 
\be 
f_{\mbox{\tiny qRM}}(\bar r_+)=0=f'_{\mbox{\tiny qRM}}(\bar r_+)\,.
\ee 
%$H(r_+)=H'(r_+)=0$. 
In contrast to the Maxwell case, the extremal black holes may not exist (see above). Moreover, 
even if they do exist,
we cannot find analytic formulae for extremal $\mu$ and $e$ since our charge parameter appears in a logarithm in the metric function, and extremal parameters can thus be obtained only numerically. Nevertheless, we can perform the following analysis similar to \cite{Climent:2024nuj} or \cite{Panella:2024sor}. Let us plot the form-function $\mu\Delta^3$, c.f. \eqref{Mf} (see also $F(M,q)$ in \cite{Climent:2024nuj}), as a function of mass $8\mathcal{G}_3M=-\kappa\Delta^2$; the result is plotted in Fig.~\ref{branchesRM}. Here, branches with negative mass correspond to previously charged defects dressed in a horizon by quantum backreaction. Note also the thin black curve for which no extremal black holes exist, indicating the regime of Schwarzschild-like black holes. The Maxwell and RegMax cases are compared in Fig.~\ref{branchescompare}. As indicated in this figure, the Reissner--Nordstr{\"o}m-like quantum RegMax black holes can also exist in the `beyond extremal' Maxwell regime.  

\begin{figure}[H]
\begin{center}
    \includegraphics[scale=0.82]{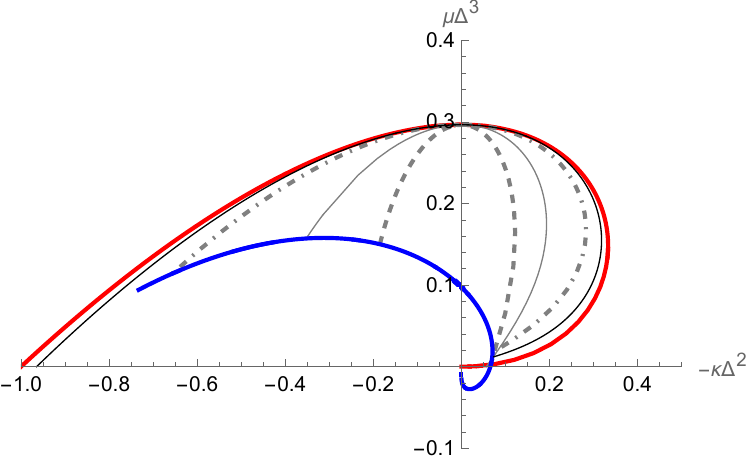}
	\caption{{\textbf{Branches of RegMax quantum black holes.} Form-function $\mu\Delta^3$ is parametrically plotted as a function of mass $M\propto-\kappa \Delta^2$. The red curve corresponds to neutral quantum black hole, the gray curves represent non-extremal charged black holes with $e=0.2$ (dot-dashed), $e=0.5$ (solid thin) and $e=1$ (dashed). Thin black line corresponds to $e=0.1$; it occurs in the parameter space where only Schwarzschild-like black holes are present.  The blue curve traces out the extremal black holes. Note that extremal black holes cease to exist for some combinations of $e$ and $\alpha$ -- see discussion of marginal cases.  For all curves, $\mathcal{A}=0.2$, $\ell_3=8$ and $\alpha=0.8$; parts with $M<0$ correspond to $\kappa=1$ while positive mass corresponds to $\kappa=-1$. }}
\label{branchesRM}
\end{center}
\end{figure}

\begin{figure}[H]
\begin{center}
    \includegraphics[scale=0.67]{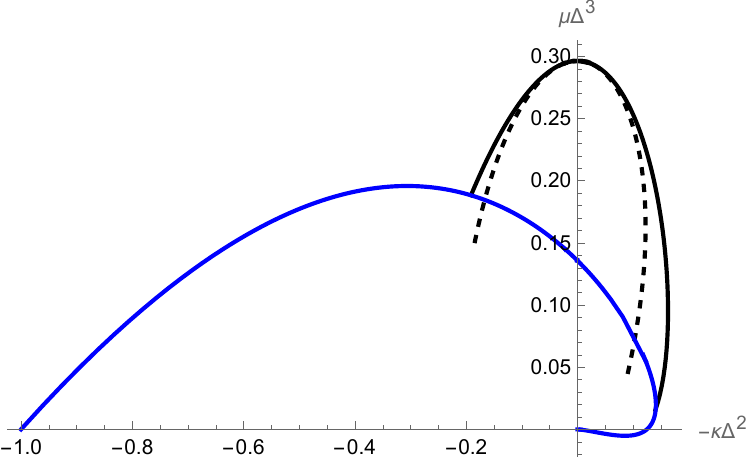}
	\caption{{\textbf{Comparison of Reissner--Nordstr{\"o}m-like %non-extremal 
    branches of Maxwell and RegMax quantum black holes.} Here, we compare curves for Maxwell (solid black) and RegMax (dashed black) quantum black holes with $e=1$. As above, $\mathcal{A}=0.2$, $\ell_3=8$ and $\alpha=0.8$ for the RegMax case. The blue curve traces out extremal black holes for Maxwell; obviously, quantum RegMax black holes can exist in the Maxwell `beyond extremal' regime (lower blue curve). }}\
\label{branchescompare}
\end{center}
\end{figure}

\subsection{Thermodynamics}

Let us finally turn towards the thermodynamics of the obtained quantum RegMax black holes.
We have already found the mass $M$ by identifying it with the sub-leading term of the metric function according to \eqref{qRMmass}, giving \eqref{qRMmass}. The entropy $S$ and the charge $Q$ of the  brane black hole can be identified with the bulk quantities,  counted twice since we have a  double-copy of the spacetime. The temperature $T$ and the chemical potential $\phi$ are then determined from \eqref{qRegMax solution} via standard means. This yields
\ba
M&=&-\frac{1}{8 {\cal G}_3}\kappa\Delta^2\,,\quad 
S=2\frac{A_4}{4G_4}=\frac{2}{4G_4}\int^{2\pi \Delta}_0\!\! \dd \varphi \int_0^{x_1} \!\!\frac{r_+^2}{(1+\mathcal{A}r_+x)^2}\dd x=\frac{\pi  r_+^2 x_1 \Delta}{G_4(1+\mathcal{A} r_+ x_1)}\,,\nonumber\\
T&=&\frac{{f}'_{\mbox{\tiny qRM}}(\bar r_+)}{4\pi}\,,\quad 
Q=\frac{2}{g_*^2}\int * D=\frac{4\pi e\alpha x_1\Delta}{g^2_*\mathcal{A}\left(\alpha-\mathcal{A}\sqrt{\frac{e}{\mathcal{A}}}\right)}\,,\quad 
\phi =\frac{\alpha e \Delta^2}{\cal A}\frac{1}{ \Delta \sqrt{\frac{e}{\mathcal{A}}}+\alpha  \Bar{r}_+}\,,
\ea
where $\Delta$ is given in \eqref{rmdelta}, $G_4=2G_3\ell_4$, $\mathcal{G}_3=\frac{\mathcal{A}}{2}G_4=\mathcal{A}\ell_4G_3$,  and we convert $\ell_4$ to $\ell_3$ by using \eqref{l4}. 
We note that all of these quantities reduce to the correct expressions in the Maxwell limit, and that the electric charge again vanishes in the limit of zero backreaction, meaning that the charge on the brane is a purely quantum effect.\footnote{In the case of the charge and the chemical potential, we can identify a difference in the numerical factors between the Maxwell and RegMax results, which is caused by the fact that, in the Maxwell case, $\ell_*$ is explicitly part of the electric potential $A$, while in the RegMax case, we have  absorbed it into the definition of the electromagnetic Lagrangian.}

Similar to the Maxwell case, we want to analytically express these quantities in terms of three  independent parameters (reflecting the dependence of the hole on its mass, electric charge, and the brane tension). This is again achieved by introducing the following variables:   
\be
z=\frac{\ell_3}{r_+x_1}\,,\quad \nu=\frac{1}{\mathcal{A}\ell_3}\,,\quad \gamma=ex_1^2\,,
\ee
which yields:\footnote{We can obtain $x_1^2$ from $f_{\mbox{\tiny qRM}}(\bar r_+)=0$ where we plug in the event horizon radius $\bar r_+=\frac{\ell_3}{x_1z}$. To  express $\mu$ in terms of $x_1$, we solve $G(x_1)=0$.}
\ba
x^2_1&=&\frac{1}{\kappa z^2(1+z\nu)}\bigg(4 \alpha ^4 \ell_3^2 \nu ^3 z^3 \log \Big(1-\frac{\gamma }{\alpha  \sqrt{\gamma  \ell_3 \nu }}\Big)+4 \alpha ^4 \ell_3^2 \log \Big(\frac{\gamma  \nu  z}{\alpha  \sqrt{\gamma  \ell_3 \nu }}+1\Big)\nonumber\\
&& +4 \alpha ^3 \ell_3 \nu ^2 z^3 \sqrt{\gamma  \ell_3 \nu }+2 \alpha ^2 \gamma  \ell_3 \nu ^2 z^3+2 \alpha ^2 \gamma  \ell_3 \nu  z^2-4 \alpha ^3 \ell_3 z \sqrt{\gamma  \ell_3 \nu }+\nu  z^3-1\bigg)\,,\nonumber\\
r_+^2&=&\frac{\ell^2_3}{x_1^2z^2}\,.
\ea
Plugging these expressions into the above formulas then yields thermodynamic quantities to all orders in $1/\alpha$. However, since such expressions 
are quite unwieldy, we do not present them here and only state their expansion to the order of $\frac{1}{\alpha}$. They read
\ba
M&=&\frac{\sqrt{\nu ^2+1} z^2 (\nu  z+1) \left(\nu  z^3 \left(\gamma ^2 (\nu  z+1)-1\right)+1\right)}{2 G_3 \left(z^2 \left((\gamma +\gamma  \nu  z)^2+2 \nu  z+3\right)+1\right)^2}+2 \gamma ^3 \sqrt{\nu ^2+1} z^4 (\nu  z+1)^2\nonumber\\
 && \times\frac{ \left(\nu  z \left(z \left(\nu +z \left((\nu  z-3) (\gamma +\gamma  \nu  z)^2+\nu  z (1-4 \nu  z)+7\right)\right)-1\right)-4\right)}{5 \alpha  G_3 \sqrt{\gamma  \ell_3 \nu } \left(z^2 \left((\gamma +\gamma  \nu  z)^2+2 \nu  z+3\right)+1\right)^3}\,,\nonumber\\
S&=&\frac{z \pi  \ell_3 \sqrt{\nu ^2+1}}{G_3 \left(\gamma ^2 \nu ^2 z^4+2 \gamma ^2 \nu  z^3+2 \nu  z^3+\gamma ^2 z^2+3 z^2+1\right)}\nonumber\\
&& +\frac{4 \pi  \gamma ^3 \ell_3 \sqrt{\nu ^2+1} z^3 (\nu  z-2) (\nu  z+1)^2}{5 \alpha  G_3 \sqrt{\gamma  \ell_3 \nu } \left(z^2 \left((\gamma +\gamma  \nu  z)^2+2 \nu  z+3\right)+1\right)^2}\,,\nonumber\\
T&=&
-\frac{z \left(\gamma ^2 \nu ^3 z^5+2 \gamma ^2 \nu ^2 z^4+\gamma ^2 \nu  z^3-\nu  z^3-3 \nu  z-2\right)}{2 \pi  \ell_3 \left(z^2 \left((\gamma +\gamma  \nu  z)^2+2 \nu  z+3\right)+1\right)}\nonumber\\
&&+\frac{2 \left(\gamma ^4 \nu  z^3 (\nu  z+1)^2 \left(\nu  z \left(z \left(5 \nu +z \left(\gamma ^2 (\nu  z+1)^3+\nu  z (4 \nu  z+5)-5\right)\right)-5\right)-4\right)\right)}{5 \alpha  \pi  (\gamma  \ell_3 \nu )^{3/2} \left(z^2 \left((\gamma +\gamma  \nu  z)^2+2 \nu z+3\right)+1\right)^2}\,,
%=
%\left(\frac{\partial M}{\partial S}\right)_Q\,,
\nonumber\\
Q&=&\frac{16 \pi  \gamma  \sqrt{\nu ^2+1} z^2 (\nu  z+1)}{5 g_3^2 \left(z^2 \left((\gamma +\gamma  \nu  z)^2+2 \nu  z+3\right)+1\right)}\nonumber\\
&&+\frac{16\pi  \gamma ^2 \sqrt{\nu ^2+1} z^2 (\nu  z+1) \left(z^2 \left((4 \nu  z-3) (\gamma +\gamma  \nu  z)^2+5 (2 \nu  z+3)\right)+5\right)}{25 \alpha  g_3^2 \sqrt{\gamma  \ell_3 \nu } \left(z^2 \left((\gamma +\gamma  \nu  z)^2+2 \nu  z+3\right)+1\right)^2}\,,\\
\phi &=&\frac{5 g_3^2  \gamma  \nu  z^3 (\nu  z+1)}{8 \pi  G_3 \left(\gamma ^2 \nu ^2 z^4+2 \gamma ^2 \nu  z^3+2 \nu  z^3+\gamma ^2 z^2+3 z^2+1\right)}\nonumber\\
&& -\frac{\gamma ^2 g_3^2 \nu  z^4 (\nu  z+1) \left(5 \nu +\gamma ^2 \nu ^3 z^4+10 \left(\gamma ^2+1\right) \nu ^2 z^3+\left(17 \gamma ^2+15\right) \nu  z^2+8 \gamma ^2 z\right)}{8 \pi \alpha  G_3 \sqrt{\gamma  \ell_3 \nu } \left(\gamma ^2 \nu ^2 z^4+2 \left(\gamma ^2+1\right) \nu  z^3+\left(\gamma ^2+3\right) z^2+1\right)^2}\,,
%=\left(\frac{\partial M}{\partial Q}\right)_S\,,
\nonumber
\ea
plus $\mathcal{O}\left(1/{\alpha ^2 }\right)$ corrections. We have verified that the full expressions satisfy the standard first law 
\be 
\delta M=T\delta S+\phi \delta Q\, 
\ee
to all orders in $1/\alpha$. 
Here, we have varied $z$ and $\gamma$ while keeping $\nu$ and $\ell_3$ constant. Similar to the Maxwell case, this thermodynamics applies to the case of the 2-brane scenario illustrated in Fig.~\ref{twobranes}.

\section{Conclusion}\label{Sec5}

In this paper, we have coupled gravity to the RegMax non-linear electrodynamics and used standard tools of holographic renormalization and braneworld holography to construct a new class of charged quantum black holes. We obtained the respective induced gravity theory on the brane; it contains a novel 3D NLE theory, given perturbatively by \eqref{IEMmain}, induced from the RegMax theory in four dimensions. It remains to be seen whether such expansion can be re-summed to yield a full 3D  Lagrangian.    

An intriguing feature of the novel quantum black holes is their rich horizon structure, containing two branches with different behaviors resembling either Schwarzschild or Reissner--Nordström black holes, reminiscent of their 4-dimensional classical counterparts \cite{Hale:2023dpf}. 
Equally interesting is the study of the thermodynamics of these objects. In the Maxwell case, the thermodynamics of the AdS C-metric is (in the case of small acceleration) well understood, e.g.  \cite{Anabalon:2018ydc, Anabalon:2018qfv, Gregory:2019dtq} (see also \cite{Kim:2023ncn}), and yields the associated thermodynamics of the Maxwell-charged  quantum black hole \cite{Climent:2024nuj}.\footnote{Especially intriguing in this respect is the fact that in the case of the bulk C-metric, one has a particularly normalized  `time-like' Killing vector defining the mass of the hole \cite{Anabalon:2018ydc} that is different from the preferred time of the brane black hole in \eqref{maxwellchargedqBTZ}.    } However, in the RegMax case, the first law for the bulk black hole still remains to be understood \cite{Hale:2025veb}. It is then quite remarkable that, as shown in this paper, the first law works for the corresponding RegMax quantum BTZ black hole. The implications of this discrepancy remain to be understood.       

We would also like to emphasize that this whole analysis was possible only because RegMax theory admits a C-metric solution describing accelerated black holes. 
An  analogous study of quantum black holes in ModMax NLE (the only other theory 
currently known to possess 
 the C-metric
 solution) could be a focus of future work, as could a study comparing bulk, brane, and boundary thermodynamics of the various possible solutions.

\subsection*{Acknowledgements} We would like to thank Robie Hennigar and Pavel Krtouš for insightful discussions. 
This work was supported in part by the Natural Sciences and Engineering Research Council of Canada.
D.K. acknowledges support from the
Charles University Research Center Grant No.
UNCE24/SCI/016.
J.M. is grateful for the support of the Mitacs Globalink Research Award and the support of the University of Waterloo,  where much of this work was carried out.

\appendix

\section{Holographic renormalization of RegMax electrodynamics}
\label{counterterms}

In this appendix, 
we shall compute the RegMax electromagnetic divergent part $\mathcal{I}_{EM}$,
\eqref{IEMmain}, stated in the main text. 
To this purpose, 
we take an
%taking an 
approach similar to that in Maxwell electrodynamics \cite{Taylor:2000xw}, and employ the 
  Fefferman--Graham expansion
\ba
 &&\dd s^2=\frac{\ell_{d+1}^2}{z^2}\dd z^2+\frac{1}{z^2}\gamma_{ij}\dd x^i\dd x^j\,,\nonumber\\
 \quad  &&\gamma_{ij}=\gamma^{(0)}_{ij}+z^2\gamma^{(2)}_{ij}+z^4\gamma^{(4)}_{ij}+\dots\,,\\
 \quad &&F=F^{(0)}+zdz\wedge A^{(1)}+z^2F^{(2)}+z^2dz\wedge A^{(2)}+z^3F^{(3)}+\dots\,,\nonumber
 \label{FGexpApp}
 \ea
where $\gamma_{ij}^{(k)}, A^{(k)}$ and $F^{(k)}$ are only functions of the transversal coordinates $x^j$.
We shall further promote the RegMax Lagrangian \eqref{LRM} to any number of dimensions.\footnote{Note, however, that such a Lagrangian has distinguished properties (including the existence of the C-metric solution) only in four dimensions; see also \cite{Kubiznak:2024ijq} for proposals on how to extend RegMax theory to higher dimensions. }

The RegMax Lagrangian \eqref{LRM} in terms of $\mathcal{S}$ reads
 \be
\mathcal{L}_{RM}= \frac{\alpha ^2 \left(6 \alpha ^2 \left(\sqrt[4]{-\mathcal{S}}-\alpha \right) \log \left(1-\frac{\sqrt[4]{-\mathcal{S}}}{\alpha }\right)+3 \alpha  \sqrt[4]{-\mathcal{S}} \left(\sqrt[4]{-\mathcal{S}}-2 \alpha \right)+(-\mathcal{S})^{3/4}\right)}{\sqrt[4]{-\mathcal{S}}-\alpha }\,,
 \ee
and
\be
\frac{\partial \mathcal{L}_{RM}}{\partial \mathcal{S}}=-\frac{\alpha ^2}{2 \left(\sqrt[4]{-\mathcal{S}}-\alpha \right)^2}\,. \label{LS}
\ee
Expanding  $F^2=2\mathcal{S}$ in powers of $z$ gives ($\mathring{R}$ and $\mathring{R}_{ij}$ denoting Ricci scalar and tensor with respect to $\gamma^{(0)}$)
\be
-\mathcal{S}=-\frac{z^4F_{(0)}^2}{2}-\frac{z^6 A_{(1)}^2}{\ell_{d+1}^2}+\frac{z^6 (\dd A\cdot F^{(0)})}{2}-\frac{z^6 \ell_{d+1}^2 F_{(0)}^2 \mathring{R}}{2(d-1)(d-2)}+\frac{z^6 \ell_{d+1}^2 \mathring{R}^i_j F_{(0)}^{kj}F^{(0)}_{ki}}{d-2}\,, \label{F2exp}
\ee
and in turn
\ba
(-\mathcal{S})^{\frac{1}{4}}&=&\tfrac{\sqrt[4]{-F_{(0)}^2} z}{\sqrt[4]{2}}\nonumber\\
&&-\tfrac{z^3 \left(2 A_{(1)}^2 \left(d^2-3 d+2\right)-\left(d^2-3 d+2\right) (\dd A\cdot F^{(0)}) \ell_{d+1}^2+\ell_{d+1}^4 (F_{(0)}^2 \mathring{R}-2 (d-1) \mathring{R}^i_j F_{(0)}^{kj}F^{(0)}_{ki})\right)}{4 \left(\sqrt[4]{2} (d-2) (d-1) (-F_{(0)}^2)^{3/4} \ell_{d+1}^2\right)}\nonumber\\
&& -\tfrac{3 z^5 \left(2 A_{(1)}^2 \left(d^2-3 d+2\right)-\left(d^2-3 d+2\right) (\dd A\cdot F^{(0)}) \ell_{d+1}^2+\ell_{d+1}^4 (F_{(0)}^2 \mathring{R}-2 (d-1) \mathring{R}^i_j F_{(0)}^{kj}F^{(0)}_{ki})\right)^2}{32 \left(\sqrt[4]{2} (d-2)^2 (d-1)^2 (-F_{(0)}^2)^{7/4} \ell_{d+1}^4\right)}\nonumber\\
&&-\tfrac{7 z^7 \left(2 A_{(1)}^2 \left(d^2-3 d+2\right)-\left(d^2-3 d+2\right) (\dd A\cdot F^{(0)}) \ell_{d+1}^2+\ell_{d+1}^4 (F_{(0)}^2 \mathring{R}-2 (d-1) \mathring{R}^i_j F_{(0)}^{kj}F^{(0)}_{ki})\right)^3}{128 \left(\sqrt[4]{2} (d-2)^3 (d-1)^3 (-F_{(0)}^2)^{11/4} \ell_{d+1}^6\right)}\nonumber\\
&& +O\left(z^9\right)\,, \label{S14}\\
(-\mathcal{S})^{\frac{3}{4}}&=&\tfrac{(-(F^{(0)})^2)^{3/4} z^3}{2^{3/4}}\nonumber\\
&&-\tfrac{3 z^5 \left(2 A_{(1)}^2 \left(d^2-3 d+2\right)-\left(d^2-3 d+2\right) (\dd A^{(1)}\cdot F^{(0)}) \ell_{d+1}^2+\ell_{d+1}^4 (F_{(0)}^2 \mathring{R}-2 (d-1) \mathring{R}^i_j F_{(0)}^{kj}F^{(0)}_{ki})\right)}{4 \left(2^{3/4} (d-2) (d-1) \sqrt[4]{-(F^{(0)})^2} \ell_{d+1}^2\right)}\nonumber\\
&& -\tfrac{3 z^7 \left(2 A_{(1)}^2 \left(d^2-3 d+2\right)-\left(d^2-3 d+2\right) (\dd A^{(1)}\cdot F^{(0)}) \ell_{d+1}^2+\ell_{d+1}^4 (F_{(0)}^2 \mathring{R}-2 (d-1) \mathring{R}^i_j F_{(0)}^{kj}F^{(0)}_{ki})\right)^2}{32 \left(2^{3/4} (d-2)^2 (d-1)^2 (-((F^{(0)})^2)^{5/4} \ell_{d+1}^4\right)}\nonumber\\
&& +O\left(z^9\right)\,, \label{S34}\\
(-\mathcal{S})^{\frac{1}{2}}&=&\tfrac{\sqrt{-F_{(0)}^2} z^2}{\sqrt{2}}\nonumber\\
&&-\tfrac{z^4 \left(2 A_{(1)}^2 \left(d^2-3 d+2\right)-\left(d^2-3 d+2\right) (\dd A^{(1)}\cdot F^{(0)}) \ell_{d+1}^2+\ell_{d+1}^4 (F_{(0)}^2 \mathring{R}-2 (d-1) \mathring{R}^i_j F_{(0)}^{kj}F^{(0)}_{ki})\right)}{2 \left(\sqrt{2} (d-2) (d-1) \sqrt{-F_{(0)}^2 } \ell_{d+1}^2\right)}\nonumber\\
&&-\tfrac{z^6 \left(2 A_{(1)}^2 \left(d^2-3 d+2\right)-\left(d^2-3 d+2\right) (\dd A^{(1)}\cdot F^{(0)}) \ell_{d+1}^2+\ell_{d+1}^4 (F_{(0)}^2 \mathring{R}-2 (d-1) \mathring{R}^i_j F_{(0)}^{kj}F^{(0)}_{ki})\right)^2}{8 \left(\sqrt{2} (d-2)^2 (d-1)^2 (-F_{(0)}^2)^{3/2} \ell_{d+1}^4\right)}\nonumber\\ 
&&+O\left(z^8\right)\,. \label{S12}
\ea
The  Ricci tensor for the metric \eqref{FGexpApp} is
\ba
R_{zz}&=&-\frac{d}{z^2}-\frac{1}{2}\left(\Tr{\gamma^{-1} \gamma^{''}}-\frac{1}{z}\Tr{\gamma^{-1} \gamma^{'}}-\frac{1}{2}\Tr{\gamma^{-1} \gamma^{'} \gamma^{-1}\gamma^{'}}\right)\,, \label{Riccixx}\nonumber\\
R_{ij}&=&-\frac{d\gamma_{ij}}{\ell_{d+1}^2 z^2}-\frac{1}{\ell_{d+1}^2}\bigg(\frac{\gamma^{''}}{2}-\frac{\gamma^{'}}{2z}-\frac{\gamma^{'}\gamma^{-1}\gamma^{'}}{2}+\frac{\gamma^{'}\Tr{\gamma^{-1}\gamma^{'}}}{4}\nonumber\\
&&\qquad -\frac{\gamma\Tr{\gamma^{-1}\gamma^{'}}}{2z}-\frac{(d-2)\gamma^{'}}{2z}+R\ell_{d+1}^2\bigg)_{ij}\,, \label{Ricciij}\nonumber\\
R_{zi}&=& \frac{1}{2}\left(\gamma^{-1}\right)^{jk}\left[\nabla_i \gamma^{'}_{jk}-\nabla_k \gamma^{'}_{ij}\right]\,, \label{Riccixi}
\ea
which we will need later to evaluate traces in the metric determinant. Note that the primes denote $z$-derivatives. We can also compute the Ricci tensor and scalar directly from the equations of motion $R_{\mu\nu}-\frac{1}{2}Rg_{\mu\nu}+\Lambda_{d+1} g_{\mu\nu}=8\pi G_{d+1}T_{\mu\nu}$, obtaining
\ba
R_{\mu\nu}&=&-\frac{d g_{\mu\nu}}{\ell_{d+1}^2}+\frac{\alpha ^2 \ell_*^2 ((d-1) F_{\mu\sigma}\tensor{F}{_\nu ^\sigma}-2 g_{\mu\nu} \mathcal{S})}{2 (d-1) \left(\sqrt[4]{-\mathcal{S}}-\alpha \right)^2}\nonumber\\
&&-\frac{\alpha ^2 g_{\mu\nu} \ell_*^2 \left(-6 \alpha ^2 \sqrt[4]{-\mathcal{S}}+6 \alpha ^2 \left(\sqrt[4]{-S}-\alpha \right) \log \left(1-\frac{\sqrt[4]{-\mathcal{S}}}{\alpha }\right)+3 \alpha  \sqrt{-\mathcal{S}}+(-\mathcal{S})^{3/4}\right)}{(d-1) \left(\sqrt[4]{-\mathcal{S}}-\alpha \right)}\,,\nonumber\\
\quad \label{Riccitensapp}
\ea
and
\ba
R&=&-\frac{d (d+1)}{\ell_{d+1}^2}-\frac{\ell_*^2}{d-1}\Bigg(\frac{2 \alpha ^2 \mathcal{S}}{\left(\sqrt[4]{-\mathcal{S}}-\alpha \right)^2}\nonumber\\
&&+\frac{\alpha ^2 (d+1) \left(-6 \alpha ^2 \sqrt[4]{-\mathcal{S}}+6 \alpha ^2 \left(\sqrt[4]{-\mathcal{S}}-\alpha \right) \log \left(1-\frac{\sqrt[4]{-\mathcal{S}}}{\alpha }\right)+3 \alpha  \sqrt{-\mathcal{S}}+(-\mathcal{S})^{3/4}\right)}{\sqrt[4]{-\mathcal{S}}-\alpha }\Bigg)\,.\nonumber\\
\quad \label{Ricciapp}
\ea
The metric determinant  is
\be
\sqrt{-g}= \sqrt{\frac{\ell_{d+1}^2}{z^2}\left(\frac{1}{z^2}\right)^d}\sqrt{\gamma}=\frac{\ell_{d+1}}{z^{d+1}}\sqrt{\gamma}\,.
\ee
We can write $\gamma_{ij}=\gamma^{(0)}_{ij}+q_{ij}$ where $q_{ij}=z^2\gamma^{(2)}_{ij}+z^4\gamma^{(4)}_{ij}+z^6\gamma^{(6)}_{ij}+\dots$. It follows that
\be
\sqrt{\gamma}=\sqrt{\gamma^{(0)}}\left(1+\frac{1}{2}q^i_i-\frac{1}{4}q^i_j q^j_i+\frac{1}{8}q^i_i q^j_j+\frac{1}{6}q^i_j q^j_k q^k_i-\frac{1}{8}q^k_k q^i_j q^j_i+\frac{1}{48}q^i_i q^j_j q^k_k+\dots\right)\,.
\ee
From this we can expand the various scalar quantities
\ba
q^i_i &=& \gamma^{ij}_{(0)}\left(z^2\gamma^{(2)}_{ij}+z^4\gamma^{(4)}_{ij}+z^6\gamma^{(6)}_{ij}+\dots\right)\nonumber\\
&=&z^2\Tr{\gamma^{(2)}}+z^4\Tr{\gamma^{(4)}}+z^6\Tr{\gamma^{(6)}}+\dots,\nonumber\\
q^i_i q^j_j &=& z^4\Big(\Tr{\gamma^{(2)}}\Big)^2+2z^6\Tr{\gamma^{(2)}}\Tr{\gamma^{(4)}}+\dots\,,\nonumber\\
q^i_j q^j_i &=& z^4\Tr{\gamma_{(2)}^2}+2z^6\Tr{\gamma^{(2)}\gamma^{(4)}}+\dots\,,\\
q^i_i q^j_j q^k_k &=& z^6\Big(\Tr{\gamma^{(0)}}\Big)^3 +\dots\,,\nonumber\\
q^k_k q^i_j q^j_i &=& z^6 \Tr{\gamma^{(2)}}\Tr{\gamma_{(2)}^2} +\dots\,,\nonumber\\
q^i_j q^j_k q^k_i &=& z^6\Tr{\gamma_{(2)}^3}+\dots\,,\nonumber
\ea
where contractions in traces are realized with $\gamma_{(0)}^{-1}$, e.g. 
\be \Tr{\gamma^{(4)}}=\Tr{\gamma^{-1}_{(0)}\gamma^{(4)}}=(\gamma_{(0)}^{-1})^{ij}\gamma^{(4)}_{ij}\,,
\ee 
and so on.
In summary,
\ba
\sqrt{\gamma}=\sqrt{\gamma_{(0)}}&&\Bigg[ 1+\frac{z^2}{2}\Tr{\gamma^{(2)}}+\frac{z^4}{2}\Tr{\gamma^{(4)}}-\frac{z^4}{4}\Tr{\gamma^2_{(2)}}+\frac{z^4}{8}\bigg(\Tr{\gamma^{(2)}}\bigg)^2\nonumber\\
&&\quad +\frac{z^6}{2}\Tr{\gamma^{(6)}}-\frac{z^6}{2}\Tr{\gamma^{(2)}\gamma^{(4)}}+\frac{z^6}{4}\Tr{\gamma^{(2)}}\Tr{\gamma^{(4)}}\nonumber\\
&&\quad +\frac{z^6}{48}\Big(\Tr{\gamma^{(2)}}\Big)^3-\frac{z^6}{8}\Tr{\gamma^{(2)}}\Tr{\gamma_{(2)}^2}+\frac{z^6}{6}\Tr{\gamma_{(2)}^3}\Bigg]\,. \label{detexp}
\ea
Now, let us compute $\gamma^{(2)}$ in terms of $\gamma^{(0)}$.
We obtain 
\be
\gamma^{(2)}_{ij}=\frac{\ell_{d+1}^2 \mathring{R}_{ij}}{(d-2)}-\frac{\ell_{d+1}^2 \mathring{R}\gamma^{(0)}_{ij}}{2(d-2)(d-1)}\,,
\ee
from inserting the
expansion \eqref{FGexpApp} into the field equations. We then have
\ba
\Tr{\gamma^{(2)}}&=&\gamma_{(0)}^{ij}\gamma^{(2)}_{ij}=\frac{\ell_{d+1}^2 \mathring{R}}{d-2}-\frac{d \ell_{d+1}^2 \mathring{R}}{2(d-2)(d-1)}=\frac{\ell_{d+1}^2 \mathring{R}}{2(d-1)}\,,\nonumber\\
\left(\Tr{\gamma^{(2)}}\right)^2&=&\frac{\ell_{d+1}^4 \mathring{R}^2}{4(d-1)^2}\,,\\
\Tr{(\gamma^{(2)})^2}&=&\gamma_{(0)}^{ik}\gamma_{(0)}^{jl}\left(\gamma^{(2)}_{kj}\gamma_{li}^{(2)}\right)=\frac{\ell_{d+1}^4 \mathring{R}_{ij}\mathring{R}^{ij}}{(d-2)^2}-\frac{\ell_{d+1}^4 \mathring{R}^2(3d-4)}{4(d-2)^2(d-1)^2}\,.\nonumber
\ea
We can further express $\Tr{\gamma^{(4)}}$ and $\Tr{\gamma^{(6)}}$ in terms of lower-order traces, which can be done using the equations of motion. To find $\Tr{\gamma^{(4)}}$, we take terms proportional to $z^2$ in Eq.~\eqref{Riccixx} for $R_{zz}$ and compare them to $z^2$-proportional terms in Eq.~\eqref{Riccitensapp} for RegMax $R_{zz}$. This gives 
\be
\Tr{\gamma^{(4)}}=\frac{1}{4}\Tr{(\gamma^{(2)})^2}+ \frac{\ell_*^2\ell_{d+1}^2 (F^{(0)})^2}{16(d-1)}\,. \label{Trg4}
\ee
Similarly, for $\Tr{\gamma^{(6)}}$, we compare $z^4$-proportional terms and obtain
\ba
\Tr{\gamma^{(6)}}&=&\frac{2}{3}\Tr{\gamma^{(2)}\gamma^{(4)}}-\frac{1}{6}\Tr{(\gamma^{(2)})^3}-\frac{\ell_*^2A_{(1)}^2(d-2)}{24(d-1)}-\frac{(\dd A^{(1)}\cdot F^{(0)})\ell^2_{d+1}\ell_*^2 }{48(d-1)} \nonumber\\
&&+\frac{(F^{(0)})^2  \mathring{R}\ell_{d+1}^4\ell_*^2}{48 (d-2) (d-1)^2}-\frac{\ell^4_{d+1}\ell_*^2 \mathring{R}^i_j F_{(0)}^{kj}F^{(0)}_{ki}}{24(d-2)(d-1) } -\frac{\ell_*^2\ell^2_{d+1}(-F_{(0)}^2)^{3/2}}{12(d-1) \sqrt{2} \alpha ^2}\,.\label{Trg6}
\ea
Using the above, we obtain
\be
\gamma^{(4)}_{ij}=\gamma^{(4)}_{ij}[R]+\frac{\ell_*^2\ell_{d+1}^2 F^{(0)}_{ik}(F^{(0)})^k_j}{4(d-4)}-\frac{3\ell_*^2\ell_{d+1}^2 F_{(0)}^2\gamma^{(0)}_{ij}}{16(d-1)(d-4)}\,,
\ee
where $\gamma^{(4)}_{ij}[R]$ is the part depending only on curvature \cite{Taylor:2000xw};  we will not need it to find the electromagnetic counterterms.
The traces of the electromagnetic parts are
\ba
\Tr{\gamma^{(4)}}_{EM}&=&\gamma^{ij}_{(0)}\left(\frac{\ell_{d+1}^2 F^{(0)}_{ik}(F^{(0)})^k_j}{4(d-4)}-\frac{3\ell_{d+1}^2 F_{(0)}^2\gamma^{(0)}_{ij}}{16(d-1)(d-4)}\right)=\frac{\ell_*^2\ell_{d+1}^2F_{(0)}^2}{16(d-1)}\,,\nonumber\\
\Tr{\gamma^{(2)}}\Tr{\gamma^{(4)}}_{EM}&=&\frac{\ell_*^2\ell_{d+1}^4F_{(0)}^2\mathring{R}}{32(d-1)^2}\,,\\
\Tr{\gamma^{(2)}\gamma^{(4)}}_{EM}&=&\gamma^{ik}_{(0)}\gamma^{lj}_{(0)}\left(\frac{\mathring{R}_{kj}\ell_{d+1}^2}{d-2}-\frac{\mathring{R}\gamma^{(0)}_{kj}\ell_{d+1}^2}{2(d-2)(d-1)}\right)\nonumber\\
&&\qquad \times \left(\frac{\ell_*^2\ell_{d+1}^2 F^{(0)}_{lm}(F_{(0)})^m_i}{4(d-4)}-\frac{3\ell_*^2\ell_{d+1}^2F_{(0)}^2\gamma^{(0)}_{li}}{16(d-1)(d-4)}\right)\nonumber\\
&=&\frac{\ell_*^2\ell_{d+1}^4 \mathring{R}^i_j F_{(0)}^{kj}F^{(0)}_{ki}}{4(d-4)(d-2)}-\frac{(7d-10)\ell_*^2\ell_{d+1}^4F_{(0)}^2\mathring{R}}{32(d-1)^2(d-2)(d-4)}\,.
\ea
Inserting all this into \eqref{detexp}, we obtain 
\ba
\sqrt{\gamma}&=&\sqrt{\gamma^{(0)}}\Bigg[1+\frac{z^2\ell_{d+1}^2 \mathring{R}}{4(d-1)}+\frac{z^4\ell_{d+1}^4 \mathring{R}^2}{32(d-1)^2}-\frac{z^4\ell_{d+1}^4 \mathring{R}_{ij}\mathring{R}^{ij}}{8(d-2)^2}+\frac{z^4\ell_{d+1}^4 \mathring{R}^2(3d-4)}{32(d-2)^2(d-1)^2}\nonumber\\
&&\qquad
 +\frac{z^4\ell_*^2\ell_{d+1}^2F_{(0)}^2}{32(d-1)}-\frac{z^6\ell_*^2A_{(1)}^2(d-2)}{48(d-1)}-\frac{z^6(\dd A^{1} \cdot F^{(0)}) \ell_{d+1}^2\ell_*^2}{96(d-1)}\nonumber\\
&&\qquad-\frac{z^6\ell_{d+1}^2\ell_*^2(-F_{(0)}^2)^{3/2}}{24(d-1) \sqrt{2} \alpha ^2} +\frac{z^6 \ell_{d+1}^4\ell_*^2 F_{(0)}^2  \mathring{R}}{96 (d-2) (d-1)^2}-\frac{z^6\ell_{d+1}^4\ell_*^2 \mathring{R}^i_j F_{(0)}^{kj}F^{(0)}_{ki}}{48(d-1)(d-2)} +\cdots \Bigg]\,. \nonumber\\
\quad
\label{detgama}
\ea

Finally, we take the action \eqref{eq:RMaction} and substitute for the Ricci scalar \eqref{Ricciapp} and the cosmological constant $\Lambda_{d+1}=-\frac{d(d-1)}{2\ell^2_{d+1}}$, to obtain
\ba
\mathcal{I}_{bulk}=\frac{1}{16\pi G_{d+1}}\int \dd^d x \int_{\varepsilon}^{x_c} \dd z \frac{\ell_{d+1}}{z^{d+1}}\sqrt{\gamma}&&\Bigg(-\frac{2 d}{\ell_{d+1}^2}-\frac{2 \alpha ^3 \ell_*^2 \sqrt[4]{-S} \left(6 \alpha ^2-9 \alpha  \sqrt[4]{-S}+2 \sqrt{-S}\right)}{(d-1) \left(\sqrt[4]{-S}-\alpha \right)^2}\nonumber\\
&&\quad -\frac{12 \alpha ^4 \ell_*^2 \log \left(1-\frac{\sqrt[4]{-S}}{\alpha }\right)}{d-1} \Bigg)\,,
\ea
where we need to substitute \eqref{detgama} for the metric determinant and plug in expansions \eqref{F2exp},\eqref{S14}, and \eqref{S12} into the RegMax Lagrangian.
We also need to consider the boundary term
\be
\mathcal{I}_{surf}=\frac{1}{8\pi G}\int_{z=\varepsilon} \dd^d x\sqrt{h}K\,,
\ee
where 
\be
\quad h_{ab}=g_{ab}-n_a n_b = \frac{1}{z^2}\gamma_{ij}\dd x^i\dd x^j\,, \qquad 
\sqrt{h}=\frac{\sqrt{\gamma}}{z^d}\,,
\ee
and the normal vector to the brane is
\be
n^a=(\frac{z}{\ell_{d+1}},0,0,0)\,, \quad  n_a  =n^b g_{ab}=(\frac{\ell_{d+1}}{z},0,0,0)\,,
\quad 
n^2=n^a n^b g_{ab}=1\,. 
\ee
The extrinsic curvature  
\be
K_{ab}=\nabla_a n_b=\partial_a n_b -\Gamma^c_{ab}n_c
\ee
has 
\be
K_{ij}=\partial_i n_j -\Gamma^z_{ij}n_z=-\frac{1}{z\ell_{d+1}}\left(\frac{1}{z}\gamma_{ij}-\frac{1}{2}\partial_z\gamma_{ij}\right)\,,
\ee
as its only non-zero components, 
since the normal vector has only a $z$-component. 
The trace is
\ba
K&=&g^{ij}K_{ij}=-\frac{z^2\gamma^{ij}}{z\ell_{d+1}}\left(\frac{1}{z}\gamma_{ij}-\frac{1}{2}\partial_z\gamma_{ij}\right)=-\frac{1}{\ell_{d+1}}\left(d-\frac{z}{2}\gamma^{ij}\partial_z \gamma_{ij}\right)\nonumber\\
&=& -\frac{1}{\ell_{d+1}}\left(d-\frac{z}{2}(\gamma_{(0)}^{ij}-z^2\gamma_{(2)}^{ij}-z^4\gamma_{(4)}^{ij})(2z\gamma^{(2)}_{ij}+4z^3\gamma^{(4)}_{ij}+6z^5\gamma^{(6)}_{ij})+\dots\right)\nonumber\\
&=& -\frac{1}{\ell_{d+1}}\Big(d-z^2\Tr{\gamma^{(2)}}-2z^4\Tr{\gamma^{(4)}}-3z^6\Tr{\gamma^{(6)}}+z^4\gamma_{(2)}^{ij}\gamma_{ij}^{(2)}\nonumber\\
&&\qquad \qquad +2z^6\gamma_{(2)}^{ij}\gamma_{ij}^{(4)}+z^6\gamma_{(4)}^{ij}\gamma_{ij}^{(2)}\Big)\,.
\ea
We thus obtain   the leading order expression 
\ba
\mathcal{I}_{div}^{EM}&=&-\frac{ (d-8)\ell_*^2\ell_{d+1} \epsilon ^{4-d}}{256 \pi (d-4)  G_{d+1}}\int \dd^d x \sqrt{\gamma^{(0)}} F_{(0)}^2\nonumber\\
&&+\frac{ 3\ell_*^2 \ell_{d+1} \epsilon ^{5-d}}{40 \sqrt[4]{2} \pi  \alpha  (d-5) (d-1)G_{d+1}}\int \dd^d x \sqrt{\gamma^{(0)}}\sqrt[4]{-F_{(0)}^2} F_{(0)}^2\nonumber\\
&&+\frac{(d-12) \ell_*^2 \ell_{d+1}  \epsilon ^{6-d}}{192 \sqrt{2} \pi  \alpha ^2 (d-6) G_{d+1}}\int \dd^d x \sqrt{\gamma^{(0)}}\sqrt{-F_{(0)}^2} F_{(0)}^2+\dots\,,
\ea
along with the $d=3$
 counterterms (after multiplying by 2, since we have a double copy of the AdS spacetime) 
\ba
\mathcal{I}_{div}^{RegMax}&&=\frac{\ell_4}{8 \pi G_4}\int \dd^3 x \sqrt{-h}\left(-\frac{5\ell_*^2}{16}F^2-\frac{ 3\ell_*^2}{10 \sqrt[4]{2}  \alpha} \sqrt[4]{-F^2} F^2 +\frac{\ell_*^2}{4 \sqrt{2}  \alpha ^2}\sqrt{-F^2} F^2+\dots\right)\nonumber\\
\quad &&=\frac{1}{16 \pi G_3}\int \dd^3 x \sqrt{-h}\left(-\frac{\Tilde{\ell}^2_*}{4} F^2-\frac{6\Tilde{\ell}^2_*}{25 \sqrt[4]{2}  \alpha}\sqrt[4]{-F^2} F^2 +\frac{\Tilde{\ell}^2_*}{5 \sqrt{2}  \alpha ^2}\sqrt{-F^2} F^2+\dots\right)\,,\nonumber\\
\quad
\ea
where we used $G_4=2G_3\ell_4$, $\ell^2_*=\frac{4\Tilde{\ell}^2_*}{5}$ and we work in the regime where $\frac{1}{\mathcal{A}} \approx\ell_4$.
Finally,
\be
\mathcal{I}_{EM}=\mathcal{I}_{div}^{RegMax}+2\int \dd^3 x \sqrt{-h}A_k j^k\,.
\ee

\normalem
%\bibliographystyle{JHEP}
%\bibliography{database.bib}

\providecommand{\href}[2]{#2}\begingroup\raggedright\endgroup

\end{document}